# Tunable and giant valley-selective Hall effect in gapped bilayer graphene

**Authors:** Jianbo Yin[1,2,†,*], Cheng Tan[3,†], David Barcons-Ruiz[1], Iacopo Torre[1], Kenji Watanabe[4], Takashi Taniguchi[4], Justin C. W. Song[5], James Hone[3], Frank H. L. Koppens[1,6,*]

**Affiliations:**

[1]ICFO–Institut de Ciencies Fotoniques, The Barcelona Institute of Science and Technology, Castelldefels (Barcelona), Spain.

[2]Beijing Graphene Institute, Beijing, China

[3]Department of Mechanical Engineering, Columbia University, New York, New York 10027, USA.

[4]National Institute for Materials Science, Tsukuba, Japan

[5]Division of Physics and Applied Physics, Nanyang Technological University, 637371 Singapore, Singapore.

[6]ICREA–Institució Catalana de Recerça i Estudis Avancats, Barcelona, Spain.

[†]These authors contributed equally to this work

*Corresponding author. Email: jyin@icfo.net, frank.koppens@icfo.eu

**Abstract:**

Berry curvature is analogous to magnetic field but in momentum space and is commonly present in materials with non-trivial quantum geometry. It endows Bloch electrons with transverse anomalous velocities to produce Hall-like currents even in the absence of a magnetic field. We report the direct observation of *in situ* tunable valley-selective Hall effect (VSHE), where inversion symmetry, and thus the geometric phase of electrons, is controllable by an out-of-plane electric field. We use high-quality bilayer graphene with an intrinsic and tunable bandgap, illuminated by circularly polarized mid-infrared light and confirm that the observed Hall voltage arises from an optically-induced valley population. Compared with molybdenum disulfide ($MoS_2$), we find orders of magnitude larger VSHE, attributed to the inverse scaling of the Berry curvature with bandgap. By monitoring the valley-selective Hall conductivity, we study Berry curvature's evolution with bandgap. This *in situ* manipulation of VSHE paves the way for topological and quantum geometric opto-electronic devices, such as more robust switches and detectors.

**One-Sentence Summary:**
Bilayer graphene carries a giant photoinduced anomalous Hall conductance and can serve as quantum geometric infrared photodetector.

**Main text:**

In 1984, Berry proved that a quantum wave function acquires a geometric phase of $2N\pi$ (*N* as integer) after an adiabatic evolution along a loop in parameter space (*1*). This geometric phase was expressed as a surface integral of a vector field, coined as Berry curvature. Globally, the integral of Berry curvature over the full parameter space, representing the Chern number, defines the nature of specific topological states. Locally, Berry curvature $\Omega(\mathbf{k})$ behaves as an effective intrinsic field, analogous to magnetic field, but acting in momentum space. Berry curvature distorts the conventional semiclassical trajectories of Bloch electrons in crystals to endow them with an anomalous velocity perpendicular to an applied electric field **E** (*2–6*):

$$\mathbf{v}(\mathbf{k}) = \frac{\partial \epsilon(\mathbf{k})}{\hbar \partial \mathbf{k}} + \frac{q}{\hbar} \mathbf{E} \times \mathbf{\Omega}(\mathbf{k}) \quad (1)$$

where **v**, **k**, $\epsilon$, **E**, $q$ and $\hbar$ are electron velocity, wavevector, band energy, electric field, carrier charge and reduced Planck's constant, respectively.

Two-dimensional Dirac materials such as graphene and transition metal dichalcogenides (TMDs) are a particularly attractive platform for realizing a host of new topological and quantum geometrical responses (*7*). For instance, low-dissipation valley currents were inferred by measuring non-local signals in graphene-boron nitride (BN) superlattices and $MoS_2$ (*8–11*); under strong laser irradiation, graphene can exhibit a Floquet-driven anomalous Hall phase (*12*). No less notable, however, is gapped bilayer graphene (GBG), which can host enormous Berry curvature density at its band edges (Fig. 1B) which have been predicted to produce a giant anomalous (or valley-selective) Hall effect (*13*). Berry curvature in many materials is often thought to be fixed by its crystal configuration; by contrast, Berry curvature in bilayer graphene can be tailored externally by tuning the inter-layer potential, providing a versatile knob to study Berry curvature-driven quantum geometric phenomena. For example, in Refs. (*14–17*) non-local valley currents were reported in GBG, bilayer graphene-BN superlattice, and twisted double-bilayer graphene, respectively. However, nonlocal resistance fingerprints have also been predicted to originate from a spatially nonuniform gap profile (*11, 18*), and recent scanning gate imaging experiments found that non-local transport signals in graphene can also arise because of charge accumulation at the edges, and therefore may not necessarily have a bulk valley Hall effect origin (*19*). In contrast to the non-local measurements, direct Hall measurements by optical pumping can quantify the expected giant Hall conductivity in GBG, and trace the evolution of Berry curvature with gap and energy. Unlike in TMDs (*20–22*) direct observations of light-induced VSHE and its associated Berry curvature distribution have not been reported in gapped graphene-based systems.

**Measurement design**

In this work we directly probe the tunable VSHE in bilayer graphene with variable bandgap $E_g$ from 0 to 0.125 eV, by valley polarizing carriers with circularly polarized infrared light. The concept is shown in Fig. 1A. First, we use top and bottom gates to apply displacement fields $D_T$ and $D_B$, as further explained in the materials and methods (*23, 24*). The average of these two fields $\bar{D} = (D_B + D_T)/2$ adds energy bias between the two layer of bilayer graphene, which breaks inversion symmetry and controls $E_g$ as well as the $E_g$-dependent Berry curvature, as shown in Fig. 1B (*13*). The displacement difference $\delta D = D_B - D_T$ controls the Fermi level $E_F$. Second, we selectively excite carriers in the K (or K′) valley with a right (or left) circularly polarized mid-infrared laser (*25–29*). This valley circular dichroism arises directly from the opposite orbital moments of electrons in valley K and K′ when inversion symmetry is broken enabling circularly polarized light to preferentially induce interband transitions in either K or K′ valleys, see also a golden rule calculation in Texts 1 & 7 of the supplementary materials (SM). When we subject the valley-polarized photocarriers to an electric field **E** from source-drain bias $V_b$, the anomalous velocity due to the Berry curvature **Ω** induces a Hall voltage, as described by equation 1. We note that the valley imbalance is essential to observing a Hall voltage: Otherwise equal valley currents from the two valleys will cancel each other out, because they are driven by opposite **Ω** (*30*). Third, by independently tuning $\bar{D}$ and $\delta D$ (Fig. S1), we can quantify the expected giant valley-selective Hall conductivity $\sigma_H$ and study its dependence on $E_g$ and $E_F$ respectively. Therefore, our measurement can directly probe anomalous (geometric) currents in non-trivial quantum geometric materials with small bandgaps.

**Spatial distributions of photoinduced valley-selective Hall voltage**

A typical Hall device is shown in Fig. 1C and consists of ultraclean hexagonal BN (hBN)-encapsulated bilayer graphene, with monolayer (or few-layer) graphene as the semi-transparent top and bottom gates. The device design is optimized for probing the Hall voltage induced by optical illumination of the bulk of the device. The active area of the device is square-shaped, with graphite contacts placed at a relatively large distance to avoid spurious photovoltage generation at the contacts. The device is mounted in an optical cryostat (device temperature 33 K) with a focussed infrared light source (spot size of ~25μm), which can be scanned over the device surface while monitoring longitudinal (along **E** and noted by xx) and transverse (perpendicular to **E** and noted by yy) voltages. The photoinduced anomalous (or valley-selective) Hall voltage $V_H$ is the transverse voltage difference between left and right circularly-polarized illuminations $V_H = V_{yy}(↻) - V_{yy}(↺)$, which is probed by a lock-in amplifier, while the polarization chirality is modulated by photoelastic modulator (PEM) at ~50 kHz (Fig. S2, and materials and methods). To confirm the origin of $V_H$, we also modulate the laser intensity (on/off) and measure the photocurrent $\Delta I_{xx} = I_{xx}(\text{on}) - I_{xx}(\text{off})$ at the longitudinal terminals, as well as the transverse photovoltage $\Delta V_C = V_C(\text{on}) - V_C(\text{off})$ between transverse terminals. The spatial distributions of $V_H$, $\Delta I_{xx}$ and $\Delta V_C$ are shown in Fig. 1, D to F. The monopolar spatial features of both $V_H$ and $\Delta I_{xx}$ indicate that these signals arise from the bulk of the device; this contrasts with the bipolar spatial features of $\Delta V_C$ that we attribute to local built-in fields at graphene-electrode contacts.

**Exciton states and bandgap-dependent VSHE**

Data establishing the high quality and intrinsic bandgap of the device are presented in Fig. 2A, in which resistivity $\rho_{xx}$ is recorded as a function of $\bar{D}$ and $\delta D$ (see line traces in Fig. S4). Here, we use $\delta D \cdot \varepsilon_0 e^{-1}$ ($\varepsilon_0$ and $e$ are vacuum permittivity and elementary charge, respectively) to quantify Fermi level sweeping as it equalizes to the gate-injected free-carrier density $n_0$ at $\bar{D}=0$. However, at $\bar{D} \neq 0$, the value of $\delta D \cdot \varepsilon_0 e^{-1}$ is larger than $n_0$ because the Fermi level needs to move across the band gap to induce free carriers. The $\bar{D}$-dependent high resistivity plateau when the Fermi level is inside the gap is shown in Fig. 2A and an instrument-limited value of $10^7$ ohms at $\bar{D} > 0.8$ V nm$^{-1}$ has been reached. This high resistivity rules out other conductive channels [e.g. hopping or a topological conductive channel (*31*)] and confirms the existence of an intrinsic gap. The relation between $E_g$ and $\bar{D}$ is further extracted from temperature-dependent resistance measurements, as shown in Fig. 2B, Fig. S5 and SM Text 2. The Arrhenius plot in Fig. S5E shows thermally-activated band conduction for the full temperature range from 37 to 230 K, demonstrating that the device quality overcomes earlier-reported challenges in gapped Dirac materials that are related to impurity-mediated conduction (*14*, *20*). The high quality is also echoed by the high $E_g$-dependent mobility $\mu$ that exceeds 100,000 cm$^2$ V$^{-1}$ s$^{-1}$ at $T < 30$ K (Fig. S6 and SM Text 3); a comprehensive device characterization that includes van der Pauw measurements can be found in SM Texts 2, 3 and 9.

The effect of infrared optical illumination is first studied by probing multiple exciton peaks, visible in the $E_g$-dependent photocurrent curves displayed in Fig. 2C. For increasing gap size, multiple exciton states resonate with incident photons, giving rise to enhanced interband absorption and photocurrent peaks (*32*). The exciton states as indicated by the left dashed line are further labelled by black crosses in Fig. 2B. The optical excitation of photocarriers manifests as a typical photoconductivity feature under source-drain bias $V_b$ as indicated by the blue arrow in the $I_{xx}$-$V_b$ curve (Fig. 2D). Photoconductivity is an essential precondition for the VSHE, which clearly differs from the photovoltaic, photo-thermoelectric and bolometric effects observed in monolayer and bilayer graphene (*33*).

In the photoconductive regime, central for the study in this work, the $\Delta I_{xx}$ and $V_H$ are measured simultaneously as a function of $E_g$ and $E_F$ by tuning $\bar{D}$ and $\delta D$ independently (Fig. 2, E and F and their linecuts in Fig. S7, photon energy $h\nu = 122.9$ meV and laser intensity $P = 1.4$ μW μm$^{-2}$). The dashed triangular zone indicates the region with the Fermi level inside the gap before applying $V_b$ and illumination on the device, where $\Delta I_{xx}$ shows a relative uniform distribution until $E_g$ approaches $h\nu$ and triggers resonant absorption at $E_g \sim 120$ meV, depicted by the blue arrow in Fig. 2E and Fig. S8. We note that the abrupt decline of $\Delta I_{xx}$ outside the triangular zone is mainly caused by a dramatic decrease of $V_b$ (Fig. S9). In contrast to the observation of one resonant peak in $\Delta I_{xx}$, $V_H$ shows two peaks (indicated by two arrows in Fig. 2F). Besides one at $E_g \sim 120$ meV, an additional peak at $E_g \sim 10$ meV is formed by a rapid rise of $V_H$ when $E_g$ increases from zero, and a decline of $V_H$ when $E_g$ increases beyond 10 meV. Qualitatively, these features are expected from the inversion symmetry-breaking requirement of the VSHE, and the fast decline of Berry curvature with increasing $E_g$, as shown in Fig. 1B. A more quantitative analysis will follow later in the text. Another prominent feature is the presence of $V_H$ outside the dashed triangular zone, such as the area around coordinate ($\delta D \cdot \varepsilon_0 e^{-1}$, $E_g$) = ($-1 \times 10^{11}$ cm$^{-2}$, 20 meV) as indicated by a white arrow in Fig. 2F. This will be further discussed later in the text as well.

**Fingerprints of VSHE**

The VSHE has key fingerprints that arise from its dependence on in-plane electric field **E** and Berry curvature **Ω** (equation 1). The first is a linear $V_H$-$V_b$ relation as shown in Fig. 3A when the laser is positioned in the middle of the device. Spatial scans are shown in Fig. S10 and reveal opposite signs of $V_H$ for $V_b = \pm 40$ mV. The second fingerprint is the dependence of $V_H$ on **Ω**, which is controlled by selectively exciting the K or K′ valley. Experimentally, we change the chirality modulation (bottom arrows in Fig. 3B and Fig. S2) to exchange the initially-excited valley, which causes a flip of $V_H$ polarity (red circles in Fig. 3B and spatial scans in Fig. S3). In a control measurement with inversion symmetry preserved, that is, $\bar{D} = 0$, $V_H$ only shows small and featureless signals (black circles in Fig. 3B, see materials and methods for more discussion), consistent with the fact that the absence of a bandgap erases the Berry curvature.

Next, we present a more quantitative analysis of the photoinduced valley-selective Hall conductivity $\sigma_H$, because we can vary several parameters in-situ such as photocarrier density $n_{ph}$, Fermi level $E_F$ and the bandgap $E_g$. Here, $\sigma_H \approx \sigma_{xx} * V_H/V_{xx}$ [(*20*) and SM Text 4], with the following measured parameters under illumination: $V_H$, longitudinal conductivity $\sigma_{xx}$, and longitudinal voltage $V_{xx}$. With this, we can study the dependence of $\sigma_H$ on $n_{ph}$, with $n_{ph}$ obtained from the measured longitudinal photoconductivity $\Delta\sigma_{xx} = n_{ph} e \mu$ (see inset of Fig. 3C and SM Text 6). We find a non-linear increase of $\sigma_H$ with $n_{ph}$ (Fig. 3C). The initial fast rise agrees with theoretical predictions [(*13*) and SM Text 8] and shows a sensitive valley-selective Hall response with $n_{ph}$. The data also show that $\sigma_H$ can be tuned by several orders of magnitude by changing the light intensity within 1.4 µW µm$^{-2}$. Theoretical estimates (dashed lines) in Fig. 3C reveal that $\sigma_H$ is very sensitive to the Fermi level $E_F$ or gate-induced carrier density $n_0$. This will be discussed later in the text.

**Evolution of $\sigma_H$**

A full parameter dependence of $\sigma_H$ can be obtained by tuning $E_g$ (via $\overline{D}$) and the carrier density (via $\delta D$). As illustrated in Fig. 4A, the carrier density is proportional to $\Delta E = |E_F - E_{\text{edge}}|$, where $E_{\text{edge}}$ represents band edges of conduction or valence bands (SM Text 5). In Fig. 4B, we focus on hole doping with $\Delta E = E_{\text{edge}} - E_F$ and plot the experimental $\sigma_H$ versus $E_g$ and $\Delta E$. Our experimental data are consistent with simple theoretical estimates of intrinsic valley-selective Hall conductivity (*13*), as presented in Fig. 4C and SM Text 8. The agreement is notable given that no adjustable fitting parameters and only experiment-based $n_{ph}$, $n_0$ and $E_g$ values are used in the theoretical estimate (Fig. S11).

Notably, the measured $\sigma_H$ in Fig. 4B reaches large values of 0.04 e$^2$/h despite very weak illuminating power ($P$ = 1.4 µW/µm$^2$, $h\nu$ = 122.9 meV). This value is four orders of magnitude higher than that measured for MoS$_2$ at the same laser intensity (*20*). The large $\sigma_H$ magnitudes arise from the giant values of Berry curvature that can be found in Dirac materials with very narrow gaps such as GBG (*13*). Indeed, when $E_g$ is increased, we find that $\sigma_H$ rapidly diminishes (see Fig. 4B and its line trace as red circles in Fig. 4D) by more than three orders of magnitude. This $E_g$-induced giant tunability of $\sigma_H$, consistent with that expected from intrinsic valley-selective Hall conductivity (*13*), highlights the extreme sensitivity of Berry curvature induced opto-electronics to gate-controlled $E_g$ (see also Fig. 1B).

Another prominent feature we observed is a non-monotonic dependence of $\sigma_H$ on $\Delta E$ as shown by the red circles in Fig. 4E (a horizontal line trace of Fig. 4B). Here we vary $\Delta E$ but use a small fixed value of $E_g$. We find that $\sigma_H$ first increases by more than an order of magnitude, peaking at $\Delta E \approx$ 9 meV, and then decreases. This dependence is consistent with theoretical estimates of $\sigma_H$ based on the extracted experiment-based $n_{ph}$ and $n_0$ as shown by blue crosses in Fig. 4E (line trace of Fig. 4C). A similar non-monotonic dependence can also be found in $\sigma_H$ versus $E_g$ (red circles in Fig. 4D), displaying a peak at $E_g \approx 10$ meV; this qualitative trend is also similarly mirrored by the intrinsic valley-selective Hall conductivity (blue crosses in Fig. 4D).

The agreement between experiment and intrinsic valley-selective Hall conductivity (*13*) can enable tracking of the quantum geometric properties of bilayer graphene. Even as the simple theoretical (intrinsic) estimates produce the same trends as the observed Hall conductivity, there exist some quantitative discrepancies; see, for example, Fig. 4, B to D. Indeed, other mechanisms such as skew scattering and side jumps may also be present (*4*, *34*). Including these, in addition to other band structure effects, such as trigonal warping (*35*), may provide improved quantitative agreement with the observed $\sigma_H$.

**Summary and prospects**

The giant (0.04 e$^2$/h at $P$ = 1.4 µW µm$^{-2}$ and $h\nu$ = 122.9 meV) and highly (gate) tunable $\sigma_H$ response in our GBG devices contrasts sharply with that of the small $\sigma_H$ found in other Dirac materials (e.g., MoS$_2$) (*20*). This makes GBG a promising platform for realizing a new "quantum geometric photodetector" in the infrared and terahertz spectral range. Such a detector can use large longitudinal voltage bias to enhance the Hall photocurrent, but at the same time suffers nearly zero dark current at transverse Hall contacts, exhibiting a high signal-to-noise ratio.

**References and Notes**

1. M. V. Berry, Quantal phase factors accompanying adiabatic changes. *Proc. R. Soc. London. A. Math. Phys. Sci.* **392**, 45–57 (1984).

2. D. Xiao, M.-C. Chang, Q. Niu, Berry phase effects on electronic properties. *Rev. Mod. Phys.* **82**, 1959–2007 (2010).

3. R. Karplus, J. M. Luttinger, Hall effect in ferromagnetics. *Phys. Rev.* **95**, 1154–1160 (1954).

4. N. Nagaosa, J. Sinova, S. Onoda, A. H. MacDonald, N. P. Ong, Anomalous Hall effect. *Rev. Mod. Phys.* **82**, 1539–1592 (2010).

5. In addition to intrinsic processes, anomalous Hall effect may also be caused by extrinsic processes such as side jump and skew scattering.

6. M. C. Chang, Q. Niu, Berry phase, hyperorbits, and the Hofstadter spectrum. *Phys. Rev. Lett.* **75**, 1348–1351 (1995).

7. J. C. W. Song, P. Samutpraphoot, L. S. Levitov, Topological Bloch bands in graphene superlattices. *Proc. Natl. Acad. Sci. U. S. A.* **112**, 10879–10883 (2015).

8. R. V Gorbachev, J. C. W. Song, G. L. Yu, A. V Kretinin, F. Withers, Y. Cao, A. Mishchenko, I. V Grigorieva, K. S. Novoselov, L. S. Levitov, A. K. Geim, Detecting topological currents in graphene superlattices. *Science.* **346**, 448–451 (2014).

9. K. Komatsu, Y. Morita, E. Watanabe, D. Tsuya, K. Watanabe, T. Taniguchi, S. Moriyama, Observation of the quantum valley Hall state in ballistic graphene superlattices. *Sci. Adv.* **4**, eaaq0194 (2018).

10. Z. Wu, B. T. Zhou, X. Cai, P. Cheung, G.-B. Liu, M. Huang, J. Lin, T. Han, L. An, Y. Wang, S. Xu, G. Long, C. Cheng, K. T. Law, F. Zhang, N. Wang, Intrinsic valley Hall transport in atomically thin $MoS_2$. *Nat. Commun.* **10**, 611 (2019).

11. T. Aktor, J. H. Garcia, S. Roche, A.-P. Jauho, S. R. Power, Valley Hall effect and nonlocal resistance in locally gapped graphene. *Phys. Rev. B*. **103**, 115406 (2021).

12. J. W. McIver, B. Schulte, F.-U. Stein, T. Matsuyama, G. Jotzu, G. Meier, A. Cavalleri, Light-induced anomalous Hall effect in graphene. *Nat. Phys.* **16**, 38–41 (2020).

13. J. C. W. Song, M. A. Kats, Giant Hall Photoconductivity in Narrow-Gapped Dirac Materials. *Nano Lett.* **16**, 7346–7351 (2016).

14. Y. Shimazaki, M. Yamamoto, I. V. Borzenets, K. Watanabe, T. Taniguchi, S. Tarucha, Generation and detection of pure valley current by electrically induced Berry curvature in bilayer graphene. *Nat. Phys.* **11**, 1032–1036 (2015).

15. M. Sui, G. Chen, L. Ma, W. Y. Shan, D. Tian, K. Watanabe, T. Taniguchi, X. Jin, W. Yao, D. Xiao, Y. Zhang, Gate-tunable topological valley transport in bilayer graphene. *Nat. Phys.* **11**, 1027–1031 (2015).

16. K. Endo, K. Komatsu, T. Iwasaki, E. Watanabe, D. Tsuya, K. Watanabe, T. Taniguchi, Y. Noguchi, Y. Wakayama, Y. Morita, S. Moriyama, Topological valley currents in bilayer graphene/hexagonal boron nitride superlattices. *Appl. Phys. Lett.* **114**, 243105 (2019).

17. S. Sinha, P. C. Adak, R. S. S. Kanthi, B. L. Chittari, L. D. V. Sangani, K. Watanabe, T. Taniguchi, J. Jung, M. M. Deshmukh, Bulk valley transport and Berry curvature spreading at the edge of flat bands. *Nat. Commun.* **11**, 5548 (2020).

18. S. Roche, S. R. Power, B. K. Nikolic, J. H. Garcia, A.-P. Jauho, Have Mysterious Topological Valley Currents Been Observed in Graphene Superlattices? *J. Phys. Mater.* (2021), doi:10.1088/2515-7639/AC452A.

19. A. Aharon-Steinberg, A. Marguerite, D. J. Perello, K. Bagani, T. Holder, Y. Myasoedov, L. S. Levitov, A. K. Geim, E. Zeldov, Long-range nontopological edge currents in charge-neutral graphene. *Nature*, **593**, 528–534 (2021).

20. K. F. Mak, K. L. McGill, J. Park, P. L. McEuen, The valley Hall effect in $MoS_2$ transistors. *Science.* **344**, 1489–1492 (2014).

21. M. Onga, Y. Zhang, T. Ideue, Y. Iwasa, Exciton Hall effect in monolayer $MoS_2$. *Nat. Mater.* **16**, 1193–1197 (2017).

22. N. Ubrig, S. Jo, M. Philippi, D. Costanzo, H. Berger, A. B. Kuzmenko, A. F. Morpurgo, Microscopic Origin of the Valley Hall Effect in Transition Metal Dichalcogenides Revealed by Wavelength-Dependent Mapping. *Nano Lett.* **17**, 5719–5725 (2017).

23. Y. Zhang, T.-T. Tang, C. Girit, Z. Hao, M. C. Martin, A. Zettl, M. F. Crommie, Y. R. Shen, F. Wang, Direct observation of a widely tunable bandgap in bilayer graphene. *Nature.* **459**, 820–823 (2009).

24. J. B. Oostinga, H. B. Heersche, X. Liu, A. F. Morpurgo, L. M. K. Vandersypen, Gate-induced insulating state in bilayer graphene devices. *Nat. Mater.* **7**, 151–157 (2008).

25. W. Yao, D. Xiao, Q. Niu, Valley-dependent optoelectronics from inversion symmetry breaking. *Phys. Rev. B.* **77**, 235406 (2008).

26. H. Zeng, J. Dai, W. Yao, D. Xiao, X. Cui, Valley polarization in $MoS_2$ monolayers by optical pumping. *Nat. Nanotechnol.* **7**, 490–493 (2012).

27. K. F. Mak, K. He, J. Shan, T. F. Heinz, Control of valley polarization in monolayer $MoS_2$ by optical helicity. *Nat. Nanotechnol.* **7**, 494–498 (2012).

28. T. Cao, G. Wang, W. Han, H. Ye, C. Zhu, J. Shi, Q. Niu, P. Tan, E. Wang, B. Liu, J. Feng, Valley-selective circular dichroism of monolayer molybdenum disulphide. *Nat. Commun.* **3**, 887 (2012).

29. A. M. Jones, H. Yu, N. J. Ghimire, S. Wu, G. Aivazian, J. S. Ross, B. Zhao, J. Yan, D. G. Mandrus, D. Xiao, W. Yao, X. Xu, Optical generation of excitonic valley coherence in monolayer $WSe_2$. *Nat. Nanotechnol.* **8**, 634–638 (2013).

30. D. Xiao, W. Yao, Q. Niu, Valley-contrasting physics in graphene: Magnetic moment and topological transport. *Phys. Rev. Lett.* **99**, 236809 (2007).

31. L. Ju, Z. Shi, N. Nair, Y. Lv, C. Jin, J. Velasco, C. Ojeda-Aristizabal, H. A. Bechtel, M. C. Martin, A. Zettl, J. Analytis, F. Wang, Topological valley transport at bilayer graphene domain walls. *Nature.* **520**, 650–655 (2015).

32. L. Ju, L. Wang, T. Cao, T. Taniguchi, K. Watanabe, S. G. Louie, F. Rana, J. Park, J. Hone, F. Wang, P. L. McEuen, Tunable excitons in bilayer graphene. *Science*. **358**, 907–910 (2017).

33. N. M. Gabor, J. C. W. Song, Q. Ma, N. L. Nair, T. Taychatanapat, K. Watanabe, T. Taniguchi, L. S. Levitov, P. Jarillo-Herrero, Hot carrier-assisted intrinsic photoresponse in graphene. *Science*. **334**, 648–52 (2011).

34. T. Ando, Theory of Valley Hall Conductivity in Graphene with Gap. *J. Phys. Soc. Japan*. **84**, 114705 (2015).

35. S. Slizovskiy, E. McCann, M. Koshino, V. I. Fal'ko, Films of rhombohedral graphite as two-dimensional topological semimetals. *Commun. Phys.* **2**, 164 (2019).

36. S. Das Sarma, S. Adam, E. H. Hwang, E. Rossi, Electronic transport in two-dimensional graphene. *Rev. Mod. Phys.* **83**, 407–470 (2011).

37. K. J. Tielrooij, J. C. W. Song, S. A. Jensen, A. Centeno, A. Pesquera, A. Zurutuza Elorza, M. Bonn, L. S. Levitov, F. H. L. Koppens, Photoexcitation cascade and multiple hot-carrier generation in graphene. *Nat. Phys.* **9**, 248–252 (2013).

38. F. Zhang, A. H. MacDonald, E. J. Mele, Valley Chern numbers and boundary modes in gapped bilayer graphene. *Proc. Natl. Acad. Sci. U. S. A.* **110**, 10546–10551 (2013).

39. I. Torre, Diffusive solver: a diffusion-equations solver based on FEniCS. *arXiv:2011.04351* (2020).

**Acknowledgments:**

We thank Mathieu Massicotte, Achim Woessner, Fabien Vialla for fruitful discussions, and Matteo. Ceccanti for helping design Fig. 1A.

**Funding:** This work is supported by

European Union's Horizon 2020 research and innovation programme under grant agreement ref. 881603 (Graphene Flagship Core Project 3) (F.H.L.K);

European Research Council (ERC) TOPONANOP under grant agreement ref. 726001 (F.H.L.K);

The government of Spain [PID2019-106875GB-I00; FJC2018-037098-I; Severo Ochoa CEX2019-000910-S (MCIN/ AEI/10.13039/501100011033)] (F.H.L.K);

Fundació Cellex, Fundació Mir-Puig (F.H.L.K);

Generalitat de Catalunya (CERCA, AGAUR, SGR 1656) (F.H.L.K);

European Union's Horizon 2020 programme under the Marie Skłodowska-Curie grant agreements VHPC ref. 747927 (J.Y.);

National Natural Science Foundation of China (grant refs. 52072043 and T2188101) (J.Y.);

National Key R&D Program of China under Grant ref. 2020YFA0308900 (J.Y.)

National Science Foundation program for Emerging Frontiers in Research and Innovation (EFRI-1741660) (C.T. and J.H.);

The Ministry of Education Singapore, under its MOE AcRF Tier 3 Award MOE2018-T3-1-002 (J.C.W.S);

Nanyang Technological University start-up grant (NTU-SUG) (J.C.W.S).

**Author contributions:**
>FHLK and JY conceived the idea and designed the experiments;
>
>JY performed experiments and analysis of the results;
>
>DBR. assisted in bandgap measurement and analysis;
>
>JY and JCWS. performed the theoretical analysis together;
>
>CT and JH fabricated the devices with KW and TT. providing hBN crystals;
>
>IT performed numerical simulation for electrical transport analysis;
>
>FHLK supervised the project;
>
>JY, FHLK and JCWS. wrote the manuscript with contributions from JH and CT

**Competing interests:** The authors declare that they have no competing interests.

**Data and materials availability:** All data are available in the main text or the supplementary materials.

## Supplementary Materials

Materials and Methods

Supplementary Text

Figs. S1 to S11

References (*36–39*)

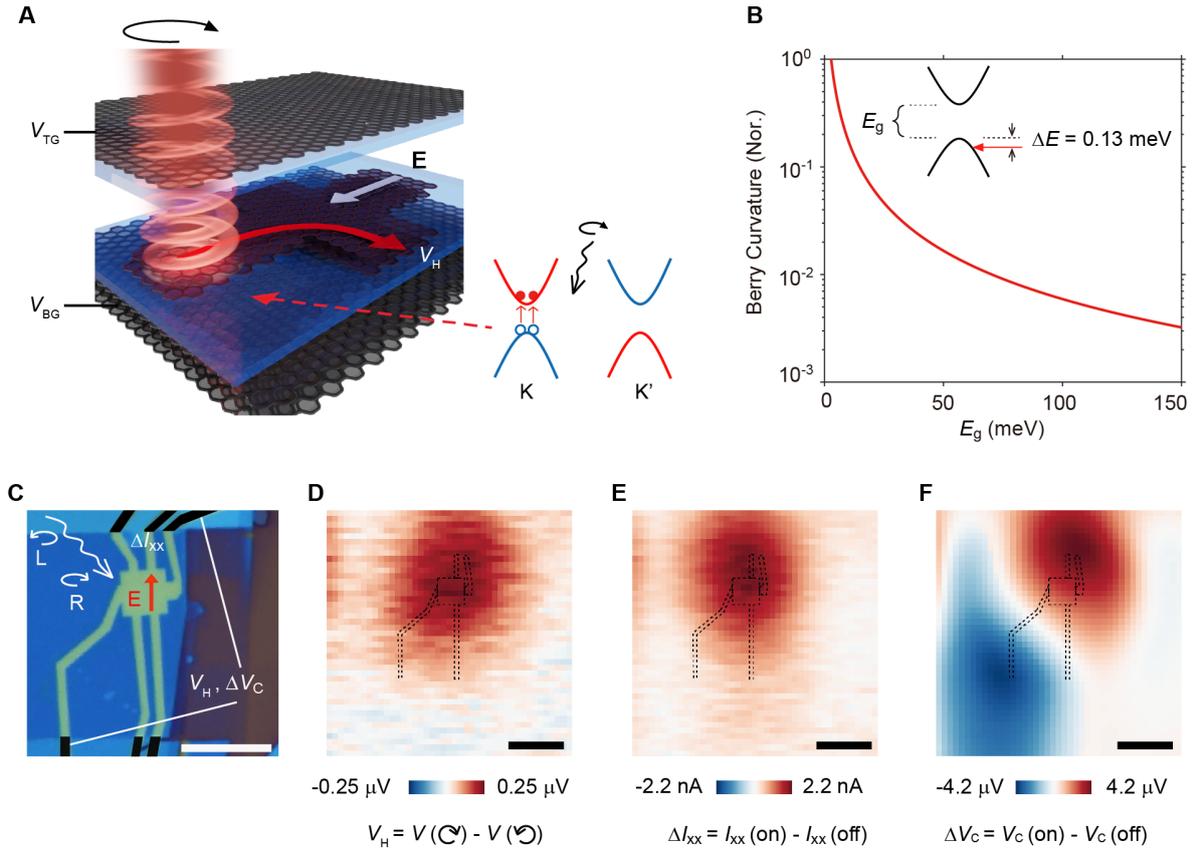

**Fig. 1. Valley-selective Hall effect in gapped bilayer graphene (GBG) driven by tunable Berry curvature.** (**A**) Schematic illustration of valley-selective Hall measurement. Photocarriers at one valley are selectively excited by circularly polarized light. These photocarriers are driven by the in-plane electric field **E** and Berry curvature $\Omega$, leading to a photoinduced anomalous Hall voltage. The blue and red colors of the illustrative band structure indicate polarities of the Berry curvature. (**B**) Fast evolution of $\Omega$ with the gap size of GBG. The curve is based on equation S16 and describes $\Omega$ values at bands with $\Delta E = 0.13$ meV away from band extremum. It is normalized by the value at $E_g = 3$ meV. Note that $\Omega$ vanishes at zero gap. (**C**) Optical image of a bilayer graphene device. To reduce photovoltage at the bilayer graphene-electrode junction, all the electrodes are made of graphite, as indicated by the artificial black color. All of the device region shown in the figure overlaps with both top and bottom gates. L, left; R, right. (**D** to **F**) Spatial distributions of valley-selective Hall voltage ($V_H$), longitudinal photocurrent ($\Delta I_{xx}$) and transverse photovoltage ($\Delta V_C$), recorded by scanning device under a focused infrared laser with photon energy $h\nu = 123.7$ meV, intensity $P = 2.2$ μW μm$^{-2}$, and beam diameter of about 25 μm. $V_H$ is collected as indicated by contacts at the furthest distance from the device bulk, while the chirality of the circularly-polarized light is modulated at 50 kHz. $\Delta I_{xx}$ and $\Delta V_C$ are collected with light intensity modulation at 177 Hz. Band gap $E_g = 35$ meV, source-drain bias $V_b = 50$ mV, gate-induced carrier density $n_0 = 0$. Dashed lines indicate device positions. Scale bars indicate 10 μm.

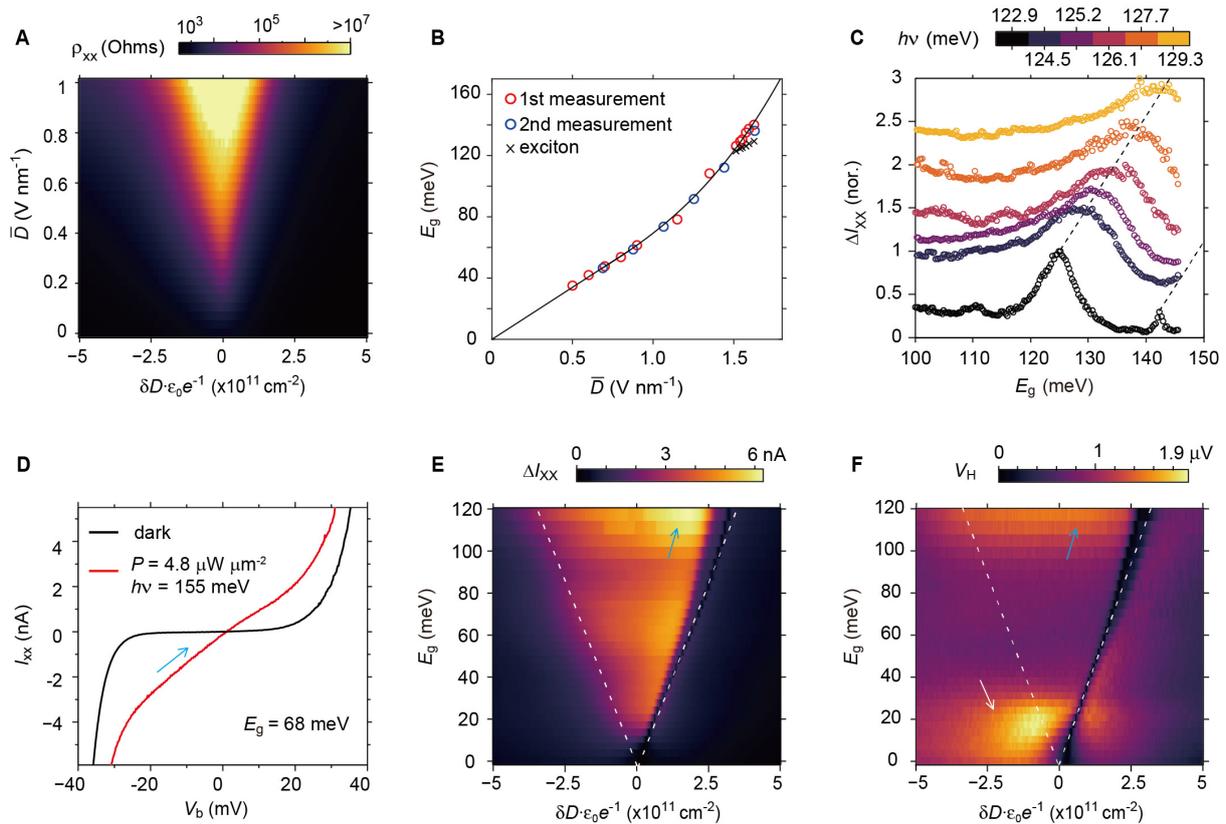

**Fig. 2. Exciton states and gap-dependent valley-selective Hall voltage ($V_H$).** (**A**) Two-dimensional map of the $\rho_{xx}$ of GBG with respect to the average ($\bar{D}$) and difference ($\delta D$) of the top and bottom displacement fields. $\bar{D}$ and $\delta D$ control band gap $E_g$ and charge carrier density (and Fermi level $E_F$) respectively. At $\bar{D} = 0$, $\delta D \cdot \varepsilon_0 e^{-1}$ is equal to the gate-induced free carrier density ($\varepsilon_0$ and $e$ are vacuum permittivity and elementary charge). $V_b = 1$ mV. (**B**) $E_g$ of GBG at different $\bar{D}$. Red and blue circles indicate data from different measurements. Each point is extracted from a series of temperature-dependent transport measurements (Fig. S5). Black crosses indicate exciton peaks, as shown by the left dashed line in panel C. (**C**) Normalized photocurrents $\Delta I_{xx}$ of GBG excited by different infrared photons, with $V_b = 0$ and the Fermi level inside the gap. Photocurrent peaks arise from excitonic resonances, which are assigned to two excited states of excitons as indicated by the dashed lines. The energy values along the left line are labelled as black crosses in panel B. Incident laser intensity ($P$) increases linearly from 0.16 to 2.6 μW μm$^{-2}$ with photon energy ($h\nu$) increasing from 122.9 to 129.3 meV, as indicated by the colour scale. The curves are shifted for clarity. (**D**) Current-bias ($I_{xx}$-$V_b$) curve of GBG in the dark and under illumination ($h\nu$ = 155.0 meV, $P$ = 4.8 μW μm$^{-2}$) with $E_g = 68$ meV and the Fermi level inside the gap before applying $V_b$ and illumination. Slope increase (blue arrow) is a typical photoconductive feature. (**E** and **F**) $\Delta I_{xx}$ and $V_H$ with respect to $E_g$ and $\delta D$ ($h\nu$ = 122.9 meV, $P$ = 1.4 μW μm$^{-2}$). The $\Delta I_{xx}$ and $V_H$ are collected by modulating light intensity and chirality of circularly-polarized light, respectively. Two dashed lines indicate the insulating region that is due to the presence of a band gap. Blue arrows indicate $\Delta I_{xx}$ and $V_H$ peaks that are due to absorption resonance. The white arrow shows and additional $V_H$ peak and the presence of $V_H$ outside the gap. The $V_b$ introduces hole doping and results in a resistance maximum around the right dashed line, along which $\Delta I_{xx}$ and hence $V_H$ decreases (see SM Text 5). $V_b$ increases from 1 to 89 mV with $E_g$ (Fig. S9A).

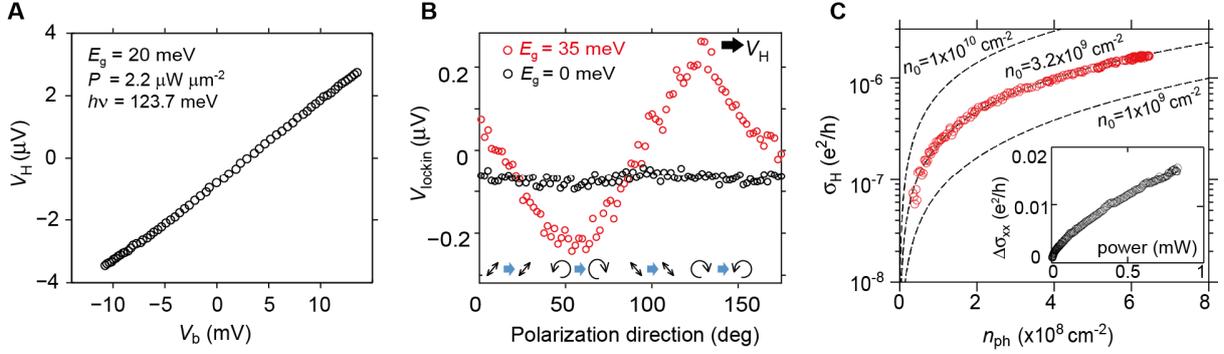

**Fig. 3. Fingerprints of VSHE in GBG.** (**A**) Linear dependence of $V_H$ on $V_b$. The inset shows band gap $E_g$, laser intensity $P$ and photon energy $h\nu$. The Fermi level is inside the gap before applying $V_b$ and illumination on the device. (**B**) Lock-in amplifier signal $V_{lockin}$ measured as a function of the polarization direction of light incident on PEM. Rotation of the polarization changes the chirality-modulation sequence from left-right to right-left (bottom arrows, see also Fig. S2). This exchanges the initially-excited valley (e.g. from K with $\Omega$ to K′ with $-\Omega$) and therefore flips $V_H$ (indicated by the black arrow, see red circles with $E_g$ at 35 meV). By contrast, linear-linear modulations at 5° and 95° do not generate valley imbalance and $V_H$. The same device without displacement field (thus without gap and measurable $\Omega$) shows a featureless curve as presented by black circles. The shift here is due to a measurement circuit instead of the device response (see materials and methods). Laser parameters are $P \sim 0.26$ μW μm$^{-2}$, $h\nu = 123.7$ meV. (**C**) Fast growth of photoinduced valley-selective Hall conductivity $\sigma_H$ with photocarrier density $n_{ph}$. Experimental results with $E_g = 122$ meV, $P = 1.4$ μW μm$^{-2}$, $h\nu = 122.9$ meV and $V_b = 70$ mV are shown by circles, which agree well with theoretical estimates of $\sigma_H$ with fitting parameter (initial carrier density) $n_0 = 3.2 \times 10^9$ cm$^{-2}$ as shown by the middle dashed line. Theoretical $\sigma_H$ with $n_0 = 1 \times 10^9$ cm$^{-2}$ and $n_0 = 1 \times 10^{10}$ cm$^{-2}$ are also shown. These three dashed lines show strong correlation between $\sigma_H$ and $n_0$. Photocarrier density $n_{ph}$ is obtained by dividing $\Delta\sigma_{xx}$ (inset) by mobility and elementary charge. For dashed theoretical lines, we treated all photocarriers as being in a single valley consistent with absorption resonance (Fig. S11C)

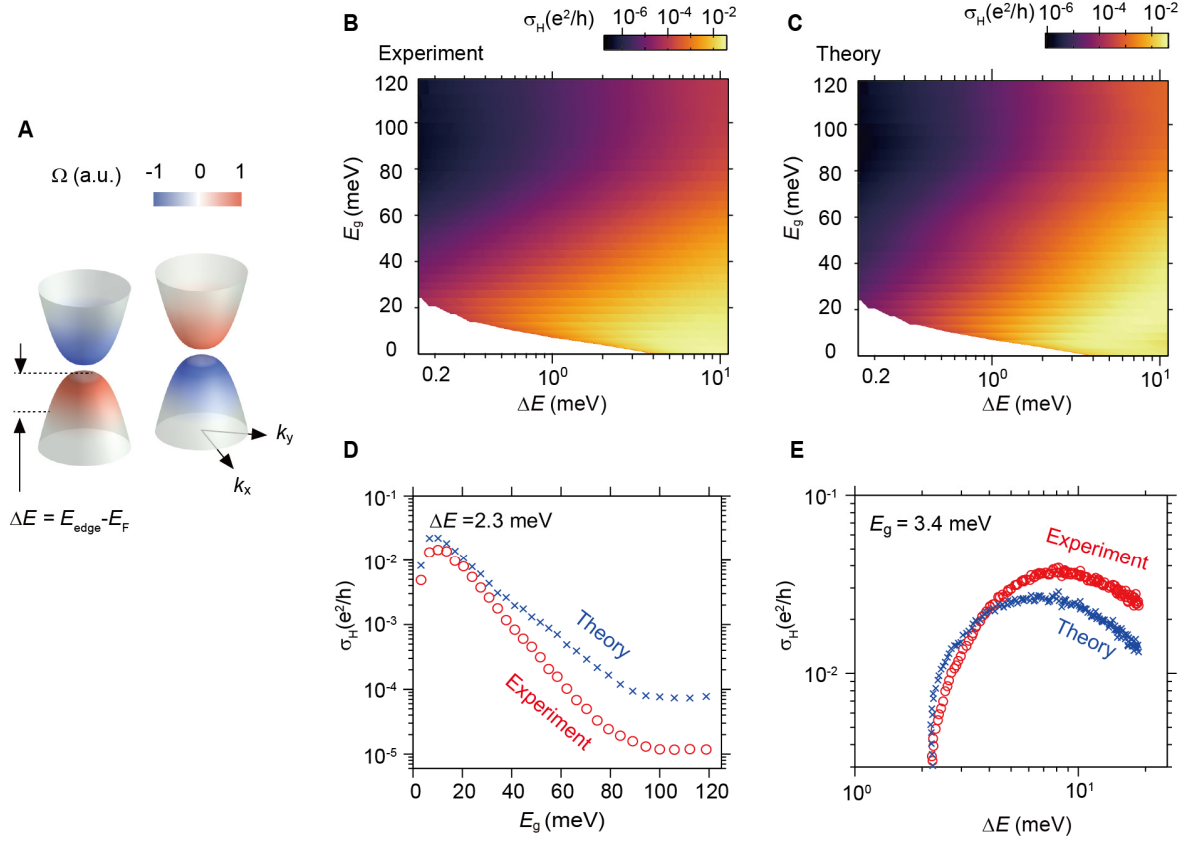

**Fig. 4. Evolution of valley-selective Hall conductivity ($\sigma_H$).** (**A**) Illustrative Berry curvature distribution in GBG. a.u., arbitrary units (**B**) Experimental dependence of $\sigma_H$ with $\Delta E$ and $E_g$. The data at the lower-left corner are not accessible because of finite carrier density at low $E_g$ induced by the bias doping effect. $h\nu =122.9$ meV and $P = 1.4$ μW μm$^{-2}$. (**C**) Theoretical estimate of intrinsic photoinduced valley-selective Hall conductivity $\sigma_H$ with $E_g$ and $\Delta E$. Here the photocarrier density $n_{ph}$ was estimated using the measured longitudinal photoconductivity and device mobility; see also SM for discussion. (**D**) Vertical line traces of panel B as circles and panel C as crosses, showing abrupt decrease of $\sigma_H$ with $E_g$. $\Delta E = 2.3$ meV. (**E**) Horizontal line traces of panel B and C, showing non-monotonic $\Delta E$-dependent $\sigma_H$. $E_g = 3.4$ meV.

# Supplementary Materials for

# Tunable and giant valley selective Hall effect in gapped bilayer graphene

Jianbo Yin[†,*], Cheng Tan[†], David Barcons-Ruiz, Iacopo Torre, Kenji Watanabe, Takashi Taniguchi, Justin C. W. Song, James Hone, Frank H. L. Koppens[*]

*Corresponding author. Email: jyin@icfo.net, frank.koppens@icfo.eu

[†]These authors contributed equally to this work

**This PDF file includes:**

    Materials and Methods
    Supplementary Text 1 to 9
    Figs. S1 to S12

## Materials and Methods

### Device fabrication

The Hall devices in our measurement consist of bilayer graphene, graphene (or thin graphite) as top gate, and graphite as bottom gates. All of these carbon layers are encapsulated and separated by boron nitride as shown in Fig. 1A and Fig. 1C. These atomically flat gates and dielectric materials guarantee a uniform gating and high breaking gate voltage. To fabricate the heterostructure in our devices, we sequentially picked up flakes of BN, graphene (or few-layer graphite) as top gate, BN, bilayer graphene as channel, graphene as electrode leads for device 1, BN, and graphite as backgate using a PPC/PDMS stamp. The PPC film with stacks were peeled off from PDMS stamp at around 110º C and left onto a Si/SiO2 substrate. The PPC was removed by vacuum annealing at T = 350 º C. To shape the device and make one-dimensional contacts, $CHF_3/O_2$ plasma was used to etch away BN, graphene and graphite at a controllable speed with a hard mask by electron-beam lithography (EBL) of a hydrogen-silsesquioxane (HSQ) resist.

We have shown two typical devices in the paper. The spatial distributions in Fig. 1, Fig. S3 and Fig. S10 are measured from device 1 with graphene as top gate as shown in Fig. 1C. All the rest data and quantitative analysis are from device 2 with very thin graphite (<10 nm) as top gate. The thicknesses of dielectric BN layers for top and bottom gates are 7.2 nm and 8.2 nm, respectively. To reduce phonon scattering, all our measurements were carried out at temperature $T$ = 33 K, unless otherwise noted.

### Displacement field and carrier density

By adjusting top and bottom gate voltages ($V_{TG}$ and $V_{BG}$), we can tune the electric displacement fields $D_T = -\varepsilon_t*(V_{TG} - V^0_{TG})/d_t$ and $D_B = \varepsilon_b(V_{BG} - V^0_{BG})/d_b$, where $\varepsilon_t$, $\varepsilon_b$, $d_t$ and $d_b$ are top/bottom gate dielectric constant and top/bottom dielectric layer thickness, and $V^0_{TG}$ ($V^0_{BG}$) are effective offset voltages of top (bottom) gates caused by initial environment-induced carrier doping (23). We note that $V^0_{TG}$ ($V^0_{BG}$) are very limited in our high-quality devices due to the encapsulation as shown in Fig. S1a.

The average of these two fields, $\bar{D} = (D_B + D_T)/2$ breaks inversion symmetry and opens a band gap in bilayer graphene, while the difference of the two fields, $\delta D = D_B - D_T$ leads to carrier doping and thus Fermi level change. In bilayer graphene with $E_g = 0$, $\delta D$ injects free charge carriers with density $n_0 = \delta D \cdot \varepsilon_0 e^{-1}$, where $\varepsilon_0$ and $e$ are vacuum permittivity and elementary charge, respectively. However, for bilayer graphene with $E_g > 0$, density of free charge carrier $n_0 < \delta D \cdot \varepsilon_0 e^{-1}$ as some of the $\delta D$-injected charge carriers are immobile and drive Fermi level across the band gap. Therefore, by adjusting $\bar{D}$ and $\delta D$ we could control band gap size and Fermi level independently. The relation between $\bar{D}$, $\delta D$ $V_{TG}$ and $V_{BG}$ is shown in Fig. S1b.

### Photoelastic modulator

To exclude a residual DC Hall voltage as shown in Fig. S2a, which is relatively large at neutralized states (usually due to impurities, device asymmetry, et. al) and drops strongly in conducting state (e.g. in the presence of relative large $V_b$ or laser illumination), we use AC modulation of light polarization and a lock-in amplifier. In the measurement the chirality of circularly-polarized light is modulated by a photoelastic modulator (PEM), which transforms periodic electrical pulses (from pulse generator) into periodic forces (compression or stretching) on ZnSe crystal. Driven by these forces, the ZnSe crystal becomes birefringent periodically and is capable of retarding the light with polarization direction parallel to the optical axis. By changing the electrical pulse intensity, one can adjust the retardation to λ/4 and turn linear polarization into circular polarization if the angle between polarization and optical axis is 45º, as shown in Fig. S2b and S2c. PEM is a resonant device with frequency $f \approx$ 50 kHz. By locking the resonant frequency with the lockin-amplifier, one can probe the modulation-related signals (eg. valley selective Hall voltage $V_H$, when modulating the chirality of circular polarization) as shown by Fig. S2d. This measurement scheme is similar to that performed for the photocurrent $\Delta I_{xx}$ and conventional photovoltage $\Delta V_C$ measurements (Fig. S2e). The only difference is that chirality is modulated by PEM with intensity unchanged for $V_H$ measurement, whereas intensity is modulated by chopper for $\Delta I_{xx}$ and $\Delta V_C$ measurement.

We note that the pulse generator of PEM works at resonant frequency and may contaminate the measurement circuit if not properly isolated. When this happens, sharing the same source with reference signal of lockin amplifier, the contamination can be picked up by the lockin amplifier and can give rise to a small constant background. This explains the offset of black circles in Fig. 3. We note that the measurements, used for quantitative analysis of valley selective Hall conductivity, are not contaminated after proper isolation.

**Supplementary Text**

Text 1. Valley selection rule

If carriers at K and K' valleys are balanced, their valley Hall voltages cancel out. Therefore, it is crucial to create carrier imbalance between two valleys in order to generate measurable valley selective Hall voltage. This means that time reversal symmetry needs to be broken (*25*). To realize this, we use a circularly-polarized infrared laser to selectively excite electrons at one valley. While this valley selective excitation has been extensively researched in transition metal dichalcogenides (TMD) (*26*). Our work demonstrates valley selective excitation in gapped bilayer graphene. This effect arises in analogy to the spin optical selection rule; instead of spin angular momentum, the valley circular dichroism arises directly from the opposite orbital moments of electrons in valley K and K' when inversion symmetry is broken enabling circularly polarized light to preferentially induce interband transitions in either K or K' valleys (*25*), see also detailed golden rule calculation in Text 7.

Text 2. Extraction of band gap.

The electrical bandgap $E_g$ of our device is determined, for each displacement field, by fitting the available temperature-dependent resistance at the charge neutrality point $R(T)$ (Fig. S5a and S5b) to a thermal activation (TA) model:

$$R(T) = R_0 * \exp\left[\frac{E_g}{2k_B T}\right] \qquad (S1)$$

Where $R_0$ is a constant, $k_B$ is Boltzmann constant. Our data shows good agreement with the TA model for $T >$ 100 K (Fig. S5c and S5d) and therefore a reliable value of the $E_g$ can be extracted at each $\bar{D}$. It is worthwhile to note that we used resistance data above 100 K to extract $E_g$, as at low temperature the resistance may be beyond the instrument range (see e.g., resistance data under high displacement field at 37 K in Fig. S5b). Arrhenius plots (natural logarithm of resistance as *y* and $T^{-1}$ as $x$) shows good linear shape as shown in Fig. S5e indicating that thermally activated conduction dominates the charge transport characteristics of our devices.

Text 3. Mobility $\mu$

The carrier mobility $\mu$ of gapped bilayer graphene is calibrated by using two geometries: van der Pauw and two terminal geometries. In the van der Pauw geometry, we monitor the mobility while tuning temperature ($T$) as shown in Fig. S6a ($E_g$ = 0 meV and $n_0$ = 1.5 × 10$^{12}$ cm$^{-2}$), S6b ($E_g$ = 0 meV and $n_0$ = −1.5 × 10$^{12}$ cm$^{-2}$) and S6c ($E_g \approx$ 40 meV and $n_0$ = −4 × 10$^{12}$ cm$^{-2}$). At temperature lower than 24~27 K as indicated by dashed lines, the gapped bilayer graphene device displays very small voltage drops across the device, indicative of a transition to a possible ballistic regime; this is consistent with the high quality of our devices. Indeed, even at higher temperatures, e.g., at 33 K, our device already shows high electron mobility of 128,500 cm$^2$ V$^{-1}$ s$^{-1}$ for $E_g$=0 meV (Fig. S6a), high hole mobility of 54,260 cm$^2$ V$^{-1}$ s$^{-1}$ for $E_g$=0 meV (Fig. S6b) and hole mobility of 37,660 cm$^2$ V$^{-1}$ s$^{-1}$ for $E_g \approx$ 40 meV (Fig. S6c). The decreasing trend of $\mu$ with $E_g$ will be further discussed below.

The two-terminal transport data are measured at $T$ = 33 K. The mobilities are fitted based on equation:

$$\rho = \frac{1}{n_0 \mu e} + \rho_c \qquad (S2)$$

where $\rho$, $\rho_c$, $n_0$ and $e$ are resistivity, constant contact resistivity, carrier density and elementary charge, respectively. The resultant electron mobility $\mu_e$ and hole mobility $\mu_h$ are shown in Fig. S6d in logarithmic scale and Fig. S6e in linear scale. In these two figures, red ($\mu_e$) and black ($\mu_h$) circles ($E_g$ range of 0 ~ 70 meV) correspond to measurement with source-drain bias $V_b$ = 1 mV as shown in Fig. 2A, while orange ($\mu_e$) and blue ($\mu_h$) circles ($E_g$ range of 0 ~ 120 meV) correspond to data from the similar measurement as shown in Fig. 2E with larger $V_b$ (see Fig. S9a). Although with different $V_b$, these two sets of data show very similar mobility results, especially at the hole side. We note that the mobility magnitudes obtained in Fig S6d and S6e (two terminals) are consistent with those in Fig. S6a, S6b and S6c (van der Pauw measurement). More importantly, the decreasing trend of mobility with gap size provides insights into transport of gapped bilayer graphene: electron mobility drops by one order of magnitude as gap increases from 0 to 120 meV, while hole mobility drops even faster by two orders of magnitude.

Text 4. Experimental valley selective Hall conductivity

Hall conductivity $\sigma_H$ in 2D system follows the equation:

$$\begin{pmatrix} j_{xx} \\ j_{yy} \end{pmatrix} = \begin{pmatrix} \sigma_{xx} & \sigma_{xy} \\ \sigma_{yx} & \sigma_{yy} \end{pmatrix} \begin{pmatrix} E_{xx} \\ E_{yy} \end{pmatrix} \qquad (S3)$$

where $j$ and $E$ are current density and electrical field, subscript xx and yy denote longitudinal and transverse directions. We have $\sigma_H = \sigma_{xy} = -\sigma_{yx}$, $\sigma_{xx} = \sigma_{yy}$, $j = I/W$, and $E = V/L$ with $I$, $W$, $V$ and $L$ as current, width, voltage and length respectively. With these relations, equation S3 gives:

$$\begin{pmatrix} I_{xx}/W_{xx} \\ I_{yy}/W_{yy} \end{pmatrix} = \begin{pmatrix} \sigma_{xx} & \sigma_H \\ -\sigma_H & \sigma_{xx} \end{pmatrix} \begin{pmatrix} V_{xx}/L_{xx} \\ V_{yy}/L_{yy} \end{pmatrix} \qquad (S4)$$

$$\frac{I_{yy}}{W_{yy}} = -\frac{\sigma_H V_{xx}}{L_{xx}} + \frac{\sigma_{xx} V_{yy}}{L_{yy}} \qquad (S5)$$

Given that (effective) length between source and drain is similar with that between two Hall bars $L_{xx} \sim L_{yy}$, and $V_{yy} = V_H$ in Hall measurement, we write:

$$I_{yy} * L_{yy}/W_{yy} = -\sigma_H V_{xx} + \sigma_{xx} V_H \qquad (S6)$$

Due to the high impedance of voltmeter ($R_{imp}$) and small voltage at Hall bars, $I_{yy}$ is usually negligible (For example, it is 1 to 4 orders of magnitude smaller than $\sigma_{xx}V_H$ for different $\bar{D}$ and $\delta D$ parameters in our measurement), which gives (see also Ref. 19):

$$\sigma_H \approx \sigma_{xx} * V_H/V_{xx} \qquad (S7)$$

The source-drain voltage $V_{xx}$ and longitudinal conductivity $\sigma_{xx, light}$ with infrared light on are shown in Fig. S9a and Fig. S9b. These two data are measured at the same time with $V_H$ in Fig. 2F, which guarantees a precise calibration of valley selective Hall conductivity $\sigma_H$. By using equation S7, we can calculate $\sigma_H$ distribution with $E_g$ and $\delta D$. In order to understand the evolution of $\sigma_H$, we have shown its distribution with $E_g$ and energy difference to band edge $\Delta E = |E_{edge} - E_F|$ as shown in Fig. 4A and 4B. The $\Delta E$ is linked to $\delta D$ via $n_0$ ($n_0$ as initial carrier density before photoexcitation, Fig. S11) as explained below.

Text 5. Link between $\delta D$, $n_0$ and $\Delta E$

In the $V_H$ measurement, the source-drain bias $V_b$ values (Fig. S9a) used are larger than that used in $\rho_{xx}$ calibration (Fig. 2A), especially when $V_b$ is inside the dashed triangle zone of Fig. S9a. This positive source-drain bias $V_b$ equivalently introduces charge carriers (e.g. holes here) into the sample. This doping can be compensated by increasing $\Delta D$ (injecting electrons). As result of the compensation, maximum resistance appears at the right border of the triangle area, around the right dashed line in Fig. 2F. When approaching this region, the resistance increases fast, leading to fast decline of $\Delta I_{xx}$ and hence fast drop of $V_H$ (see Fig. 2E & 2F). The details of the bias-induced doping are as follows.

For a neutralized gapped state, the top gate voltage $V_{TG}$ (eg. $V_{TG} > 0$) is balanced by a negative bottom gate voltage. Applying a positive source-drain bias ($V_b > 0$) weakens the gating effect of top gate (suppose $V_{TG} > 0$) by decreasing the equivalent gate voltage to $V_{TG} - V_b$. Meanwhile, it strengthens the back gate (now is negative) by increasing the amplitude of equivalent gate voltage. More negative back gate voltage and less positive top gate voltage both mean that more holes are injected to gapped bilayer graphene. This doping is inevitable if source-drain bias is applied.

The doped charge carriers elevate $\sigma_{xx,dark}$ (longitudinal conductivity before photoexcitation), which shows $10^{-7}$ to $10^{-5}$ S in the triangle zone as indicated by dashed lines in Fig. S9c. To quantify the total carrier density ($n_0$), we use equation $\sigma_{xx} = n_0 e \mu_h$ to calculate data in the hole-doped state as shown in trapezoid zone in Fig. S11a. Here, we are interested in hole doped region, and have used $\mu_h$ as shown by the blue circles in Fig. S6d, with assumption that $\mu_h$ is independent of hole carrier density for each band gap.

The resultant $n_0$ values in Fig. S11 are further used for calculating the corresponding Fermi level $E_F$. In bilayer graphene without a gap, $E_F$ follows equation (*36*):

$$E_{\mathrm{F}} = \mathrm{sign}(n_0) * 2\pi\hbar^2 n_0/(m_{\mathrm{eff}} g_s g_v) \quad (S8)$$

where effective mess $m_{\mathrm{eff}} = 0.03 m_e$ with $m_e$ as electron mass, $\hbar$ is reduced Plank constant, and $g_s = 2$ and $g_v = 2$ are spin and valley degeneracy. With the presence of a gap, we can estimate the energy difference $\Delta E$ between Fermi level $E_{\mathrm{F}}$ and band edge $E_{\mathrm{edge}}$:

$$\Delta E = |E_{\mathrm{edge}} - E_{\mathrm{F}}| \approx 2\pi\hbar^2 * |n_0|/(m_{\mathrm{eff}} g_s g_v) \quad (S9)$$

Note that we have ignored the change of band profile when $E_g$ increases, so $\Delta E$ is more precise at small gap.

With the link between $\delta D$, $n_0$ and $\Delta E$, we have focused on the hole-doped side as shown by the trapezoid in Fig. S11, and presented the corresponding experimental dependence of $\sigma_H$ on $\Delta E$ and $E_g$ in Fig. 4B.

Text 6. Photocarrier density ($n_{\mathrm{ph}}$) calculation

The total photocarrier density ($n_{\mathrm{ph}}$) including photocarriers at both valleys satisfies equation $\Delta\sigma_{xx} = n_{\mathrm{ph}} e \mu_{\mathrm{ave}}$, in which $\Delta\sigma_{xx}$ is the increase of longitudinal conductivity due to illumination as shown in Fig. S9d ($\Delta\sigma_{xx} = \sigma_{xx,\mathrm{light}} - \sigma_{xx,\mathrm{dark}}$ with $\sigma_{xx,\mathrm{light}}$ and $\sigma_{xx,\mathrm{dark}}$ as shown in Fig. S9b and S9c), and the mobility $\mu_{\mathrm{ave}}$ here is an average mobility of electron and hole: $\mu_{\mathrm{ave}} = (\mu_e + \mu_h)/2$ (with $\mu_e$ and $\mu_h$ as shown by orange and blue circles in Fig. S6d) as photocarriers include both electrons and holes. By assuming that $\mu_{\mathrm{ave}}$ only varies with $E_g$ and is independent of $n_0$, we have calculated $n_{\mathrm{ph}}$ by dividing $\Delta\sigma_{xx}$ by $\mu_{\mathrm{ave}}$ at each value of energy gap, as shown in Fig. S11b. Note that we have focused on hole-doped region here as shown by the trapezoid zone in Fig. S11b. The $n_{\mathrm{ph}}$ here includes initial photo-excited carriers and secondary hot carriers that arise from electron-electron scattering during the thermalization (*37*). The total photoexcited carrier density is highly $E_g$- and $n_0$- dependent, which explains the non-uniform distribution of $n_{\mathrm{ph}}$ in Fig. S11b.

Text 7. Valley imbalance of photocarriers

For bilayer graphene, absorption of circularly-polarized light causes valley imbalance of photocarriers $\delta n = n_{\mathrm{ph}}^K - n_{\mathrm{ph}}^{K'}$, where $n_{\mathrm{ph}}^K$ ($n_{\mathrm{ph}}^{K'}$) is photocarrier density at K (K') valley and $n_{\mathrm{ph}}^K + n_{\mathrm{ph}}^{K'} = n_{\mathrm{ph}}$. For simplicity and clarity, in what follows we use a simple two-band model for gapped bilayer graphene $H = \boldsymbol{d}(\boldsymbol{p}) \cdot \boldsymbol{\sigma}$, with (*38*)

$$\boldsymbol{d}(\boldsymbol{p}) = \left(\frac{v^2}{\gamma_1}[p_y^2 - p_x^2], \frac{2v^2}{\gamma_1} p_y p_x \zeta, \Delta\right) \quad (S10)$$

Here $\boldsymbol{\sigma} = (\sigma_x, \sigma_y, \sigma_z)$ are the Pauli matrices, $\zeta = \pm 1$ for valley K and K', $v$ is the velocity in monolayer graphene, and $\gamma_1$ is the interlayer coupling. We now turn to calculate the imbalance ratio $\delta n/n_{\mathrm{ph}}$. Using Fermi's golden rule [see e.g., Ref (*13*)], we find the rate of electron-hole pair generation rate in each valley and each spin for circularly polarized incident light is

$$W_\zeta = \frac{e^2 |E_{\mathrm{light}}|^2}{8\hbar^2 \omega} \left(\frac{2\Delta}{\hbar\omega} + \zeta d\right)^2 \quad (S11)$$

Where $d = \pm 1$ for left and right circularly polarized light, $E_{\text{light}}$ represents electrical field intensity of light, $\Delta = E_g/2$ is half band gap, $\omega$ is angular frequency of light, and $\hbar\omega \geq 2\Delta$.

With illumination of left (or right) circularly polarized light, the generation of electron-hole pairs in K and K' valleys are asymmetric, which gives the generation rate of valley imbalance (per spin) as

$$\frac{d\delta n}{dt} = 2(W_K - W_{K'}) = 2d\frac{e^2|E_{\text{light}}|^2}{\hbar^2\omega}\frac{\Delta}{\hbar\omega} \quad (S12)$$

Where the factor 2 here accounts for the generation of electrons and holes in the photoexcitation process. Similarly, the generation rate for total excited photocarrier density (per spin) follows as:

$$\frac{dn_{ph}}{dt} = 2(W_K + W_{K'}) = \frac{e^2|E_{\text{light}}|^2}{2\hbar^2\omega}\left[\left(\frac{2\Delta}{\hbar\omega}\right)^2 + 1\right] \quad (S13)$$

where we have noted $d^2 = 1$. The ratio of imbalance rate to total photocarriers rate at both valleys is:

$$\frac{d\delta n}{dt} \bigg/ \frac{dn_{ph}}{dt} = \frac{4d\left(\frac{\Delta}{\hbar\omega}\right)}{\left[\left(\frac{2\Delta}{\hbar\omega}\right)^2 + 1\right]} \quad (S14)$$

As an illustration of the effect of photon frequency and gap size on the valley imbalance, we can use Eq. (S14) to obtain a simple estimate of the valley imbalance population (for a given total photoexcited carrier density). Assuming that the relaxation of the photo-excited carriers is dominated by intra-valley processes, and writing $\Delta = E_g/2$, we estimate:

$$\frac{\delta n}{n_{ph}} = \frac{2d\left(\frac{E_g}{\hbar\omega}\right)}{\left(\frac{E_g}{\hbar\omega}\right)^2 + 1} \quad (S15)$$

For constant photon energy, such as $\hbar\omega = 122.9$ meV in our experiment, the ratio in S15 grows linearly with $E_g$ when $(E_g/\hbar\omega)^2 \ll 1$. When $E_g$ becomes larger and approaches $\hbar\omega$, $\delta n/n_{ph}$ rises sub-linearly and saturates to 1, as shown in Fig. S11c.

Text 8. Intrinsic photoinduced valley selective Hall conductivity

For the convenience of the reader, here we discuss the photoinduced Hall conductivity $\sigma_H$ for gapped bilayer graphene arising from the anomalous velocity. The Berry curvature distributions at K and K' valley of gapped bilayer graphene follows (*38*):

$$\Omega_{\pm}^{\zeta}(\mathbf{p}) = \pm\zeta\frac{2\hbar^2\Delta\gamma_1 v^4|\mathbf{p}|^2}{(v^4|\mathbf{p}|^4 + \gamma_1^2\Delta^2)^{3/2}} \quad (S16)$$

where $\pm$ denotes the conduction and valence bands, $\zeta = \pm 1$ denotes K and K' valleys, $\Delta = E_g/2$ is half of the band gap, $v$ is the velocity in monolayer graphene, $\gamma_1 = 0.38$ eV is the interlayer coupling, $\mathbf{p}$ is momentum. From equation 1 in the main text and equation $j = \sigma * E$, valley selective Hall conductivity $\sigma_H$ from the Berry curvature distribution can be written as:

$$\sigma_H = \frac{Ne^2}{h}\left[\sum_{\mathbf{p},\pm} f_\pm^K(\mathbf{p})\Omega_\pm^K(\mathbf{p}) + \sum_{\mathbf{p},\pm} f_\pm^{K'}(\mathbf{p})\Omega_\pm^{K'}(\mathbf{p})\right] \qquad (S17)$$

where N = 2 denotes spin degeneracy, $f$ is distribution function of carriers. Under photoexcitation, $f$ describes the quasi-Fermi level distribution in the conduction and valence band (at K valley as example):

$$f_\pm^K(\mathbf{p}) = [1 + e^{\beta(\varepsilon_\mathbf{p}^\pm - \mu_{\pm,K})}]^{-1} \qquad (S18)$$

Where $\pm$ denotes the function for conduction and valence bands, $\beta = 1/k_B T$ with $k_B$ and $T$ as Boltzmann constant and temperature, $\mu_{\pm,K}$ is electron/hole quasi-Fermi levels at K valley, and $\varepsilon^\pm$ is conduction/valence band dispersion. The total photoexcited carrier density is $n_{ph}$ and the initial carrier density before excitation is $n_0$ as described above.

For gapped bilayer graphene with small band gap, $\varepsilon^\pm$ can be written as:

$$\varepsilon^\pm = \pm\sqrt{v^4|\mathbf{p}|^4/\gamma_1^2 + \Delta^2} \qquad (S19)$$

Using equation S17, the carrier distributions multiplied by Berry curvatures yield:

$$\sigma_H = \frac{Ne^2}{h}\left[G_K\left(\frac{n_{ph,spin}^K}{2}\right) + G_{K'}\left(\frac{n_{ph,spin}^{K'}}{2}\right)\right] \qquad (S20)$$

$$G_\zeta(x) = \zeta\left\{1 - \left[\frac{(\tilde{n}n_1)^{1/2}}{\sqrt{x^2 + \tilde{n}n_1}} + \frac{(\tilde{n}n_1)^{1/2}}{\sqrt{(x + n_{0,sp,val})^2 + \tilde{n}n_1}}\right]\right\} \qquad (S21)$$

where $n_{ph,spin}^K$ ($n_{ph,spin}^{K'}$) are photocarrier density at K (K') valley per spin, $n_{0,sp,val}$ is initial carrier density per valley and per spin, $G_\zeta$ with $\zeta = \pm 1$ denote contributions of the K and K' valleys respectively, and $\tilde{n} = \Delta^2/(16\pi v^2\hbar^2)$ depends on gap size $E_g = 2\Delta$, $v$, and $\hbar$, $n_1 = \gamma_1^2/(4\pi v^2\hbar^2)$. This analysis is consistent with Ref. (*13*).

As seen from equations above, $\sigma_H$ depends on four variables, including $\Delta$ (or $E_g/2$), $n_{ph,spin}^K$, $n_{ph,spin}^{K'}$ and $n_{0,sp,val}$. At each experimental point with coordinate ($\delta D \cdot \varepsilon_0 e^{-1}$, $E_g$) as shown in Fig. S9, these four variables can be extracted to calculate the theoretical estimate in Fig. 4C via equations S20 and S21. In particular, at each coordinate ($\delta D \cdot \varepsilon_0 e^{-1}$, $E_g$), $n_0$ and $n_{ph}$ values were extracted as shown in Fig. S11a and S11b as explained before. $n_0$ is valley- and spin-independent and therefore is distributed equally into valleys and spins ($n_{0,sp,val} = n_0/4$). $n_{ph}$ is spin-independent but valley-dependent. It is distributed into two valleys based on relation in Fig. S11c, and then distributed equally into two spins ($n_{ph} = n_{ph}^K + n_{ph}^{K'}$, where $n_{ph}^K = 2 * n_{ph,spin}^K$ and $n_{ph}^{K'} = 2 * n_{ph,spin}^{K'}$). With all these values ($E_g$, $n_{ph,spin}^K$, $n_{ph,spin}^{K'}$ and $n_{0,sp,val}$) from experiment, the theoretical $\sigma_H$ values are calculated based on equations S20 and S21. By linking $\delta D \cdot \varepsilon_0 e^{-1}$ - $E_g$ axes to $\Delta E$ - $E_g$ axis in the same fashion as shown in Fig. 4B, we display theoretical estimates for the (intrinsic) valley selective Hall conductivity $\sigma_H$ values in Fig. 4C.

We note that Fig. 4B and 4C are supported by the same underlying $n_{ph}$ matrix (Fig. S11b), that varies at different coordinates. It is remarkable that the theoretical plot of Fig. 4C closely mirrors that found in the experiment in Fig. 4B. This close concordance between the two provides evidence that Berry curvature is essential for the $\sigma_H$ values for our devices. We further note that when $E_g$ approaches photon energy, the magnitude of $n_{ph}$ increases likely due to absorption resonance (Fig. S8). This explains the rather flat trend in $\sigma_H$ when $E_g > 100$ meV as shown in Fig. 4D: The increase of $n_{ph}$ elevates the magnitude of $\sigma_H$ (Fig. 3C), competes with the $E_g$ - dependent declining trend in $\sigma_H$ and hence results in flatness.

Text 9. Van der Pauw (vdPw) resistance $R_{34}$ vs. longitudinal resistivity $\rho_{xx}$

As shown in Fig. S12, the vdPw resistance $R_{34}$, which is obtained by dividing the measured voltage at terminal 3 and 4 by the current injected from 1 to 2, is non-zero. Although it is challenging to differentiate ohmic contribution from non-ohmic due to small $l/w$ ratio, it is useful to examine the scaling of $R_{34}$ with $\rho_{xx}$. We find a parameter regime that scales as $R_{34} \propto \rho^3_{xx}$. Note that conventional ohmic contributions are expected to scale linearly with $\rho_{xx}$. This cubic scaling at low $\rho_{xx}$ suggests that there is at least one regime that displays a non-ohmic contribution to $R_{34}$. We note that these results are only indicative, given the very small and non-ideal l/w ratio (about 0.5) of our device (which was designed for photoinduced VSHE and not for nonlocal measurements).

It is instructive to compare the conditions under which photo-induced anomalous Hall currents (from the VSHE) manifest with that of the non-ohmic contributions to vdPw measurements (as shown in Fig. S12). As we now explain, photo-induced anomalous Hall currents occur under different experimental conditions as compared with the vdPw measurements (and by extension, photo-induced anomalous Hall currents are also distinct from that of nonlocal resistance measurements in long Hall bars in GBG (*14*, *15*). Importantly, photo-induced anomalous Hall currents arise when GBG is pushed far out-of-equilibrium (by the incident light) distinct to that of the close-to-equilibrium vdPw measurements. Furthermore, photo-induced anomalous Hall currents only manifest when time-reversal symmetry is explicitly broken by the incident circularly polarized light. This contrasts with the time-reversal-symmetry of the non-irradiated GBG system used in vdPW measurements.

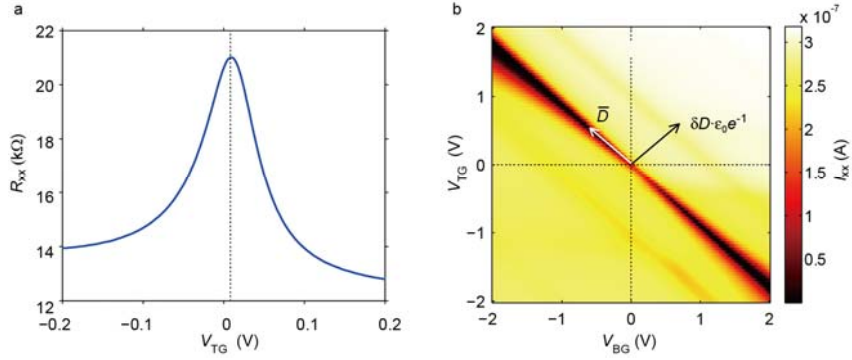

**Fig. S1. Transport measurement** (a) Top gate-dependent resistance when bottom gate voltage is 0 V. The offset of maximum resistance is around 0.01 V. This small value implies a high quality of our device. (b) Drain current with respect to top and bottom gate voltage. The source-drain bias is at 1 mV with frequency at 177 Hz. The arrows indicate the axis directions of $\bar{D}$ and $\delta D \cdot \varepsilon_0 e^{-1}$. We use $\delta D \cdot \varepsilon_0 e^{-1}$ ($\varepsilon_0$ and $e$ are vacuum permittivity and elementary charge) as axis instead of $\delta D$, because it equals to free carrier density injected by gates when $\bar{D} = 0$. However, when $\bar{D} \neq 0$, $\delta D \cdot \varepsilon_0 e^{-1}$ is larger than the free carrier density, as some of the $\delta D$-injected charge carriers are immobile and drive Fermi level across the band gap. The thicknesses of top and bottom dielectric BN layers are 7.2 nm and 8.2 nm, respectively.

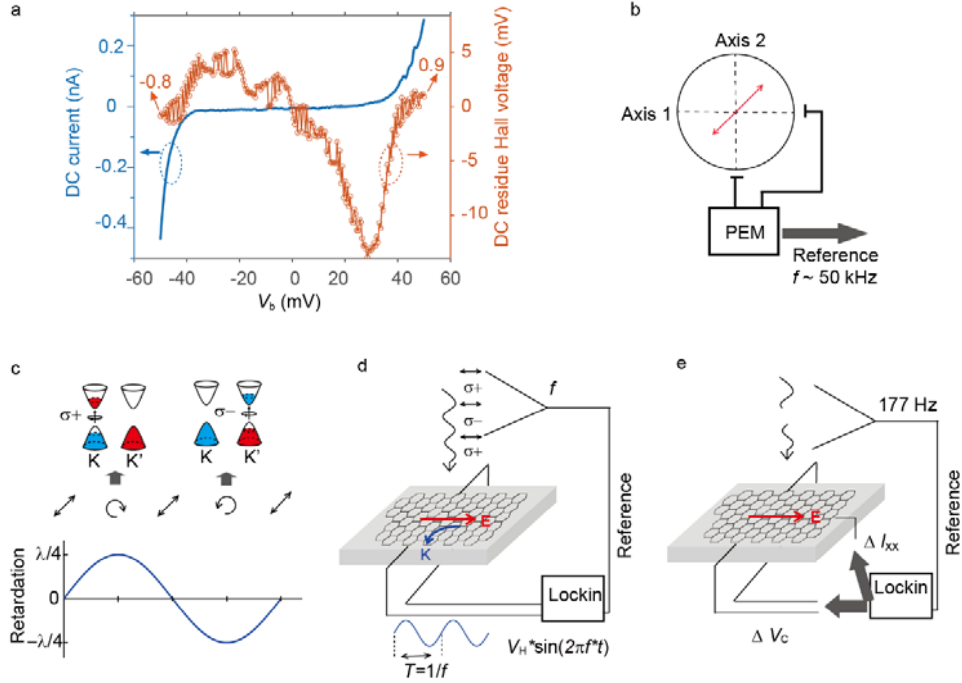

**Fig. S2. Measurement mechanism.** (a) Typical Remaining DC Hall voltage. Blue curve with left y axis is current-bias (I-V) curve of gapped bilayer graphene with band gap $E_g \sim 130$ meV. Orange circles with right y axis are remaining DC Hall voltages. The remaining Hall voltage suggests that DC measurements are not suitable for valley selective Hall measurements. (b) Polarization direction of incident laser with respect to optical axes of photoelastic modulator (PEM). The polarization direction is indicated by red arrow and has 45 degrees with respect to axes of PEM. The PEM changes the retardation between axis 1 and 2 with reference $f \approx 50$ kHz. (c) Chirality modulation by PEM. At retardation of $\lambda/4$ and $-\lambda/4$, the output beam shows right and left circular polarization, respectively. They selectively excite photocarriers at K or K' valleys. The change of chirality has frequency of $f \approx 50$ kHz. The light intensity is not changed during modulation (d). Illustration of our measurement setup. Photons with right circular polarization give rise to positive valley selective Hall voltage via K valley excitation, while photons with left circular polarization introduce negative valley selective Hall voltage. Modulating the chirality of circular polarization gives rise to valley selective Hall voltage at the same frequency of about 50 kHz. (e) The measurement of longitudinal photocurrent $\Delta I_{xx}$ or conventional photovoltage ($\Delta V_C$) at Hall bar contacts. The intensity modulation is achieved by optical chopper at 177 Hz. Lockin amplifier was used to measure $\Delta I_{xx}$ or $\Delta V_C$ with reference signal from chopper.

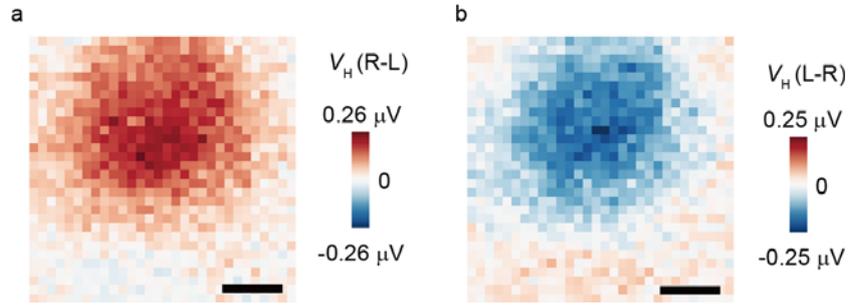

**Fig. S3. Valley selective Hall voltages $V_H$ for opposite modulation sequence.** (a) Spatial map of $V_H$ at modulating sequence of right to left (R-L) circular polarization. The modulating frequency is about 50 kHz. (b) The $V_H$ map that is similar with panel a but at opposite chirality modulating sequence from left to right (L-R). The $V_H$ shows opposite polarity to panel a, which agrees with equation 1 in main text. All the measurements were carried out at $E_g \approx 35$ meV with source-drain bias $V_b = 100$ mV.

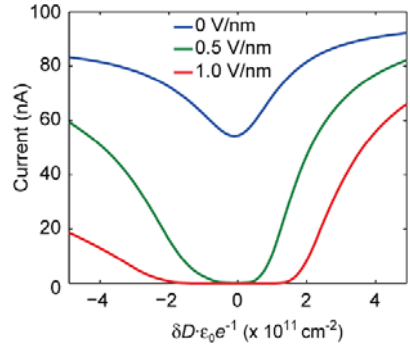

**Fig. S4. Transfer curve of gapped bilayer graphene at different electrical displacement field $\bar{D}$.** The source-drain bias $V_b$ is 1 mV at 177 Hz. The signal is firstly pre-amplified and then measured by lockin amplifier. At $\bar{D} = 0.5$ V nm$^{-1}$ and $\bar{D} = 1.0$ V nm$^{-1}$, plateaus of nearly zero drain current indicate gap-induced insulating states.

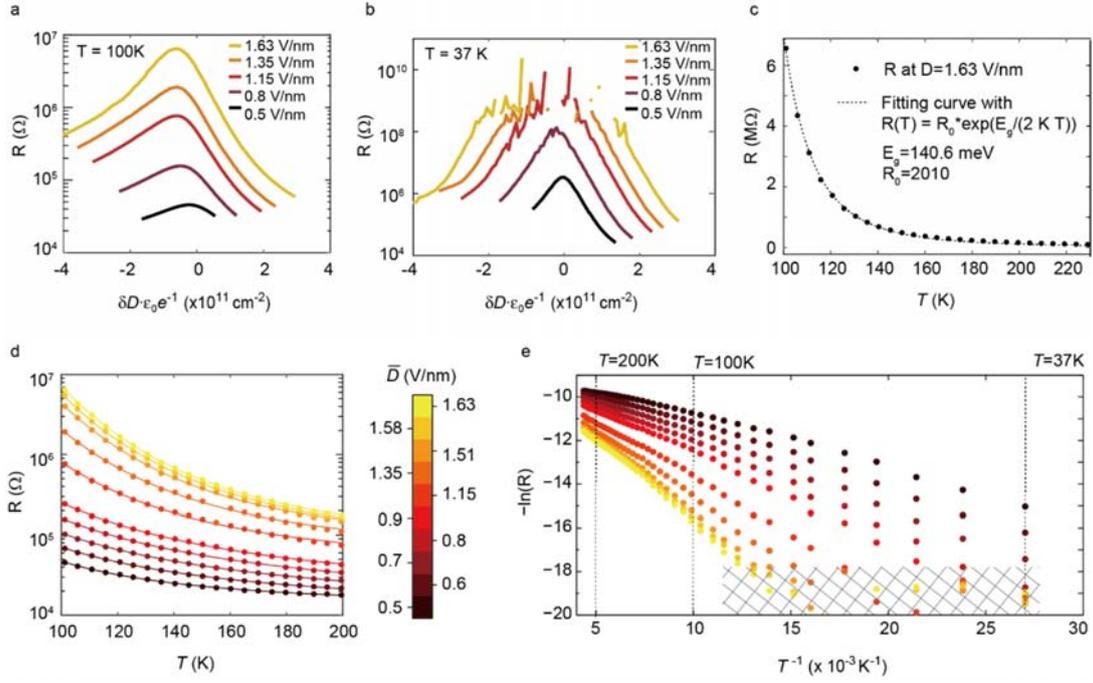

**Fig. S5. Temperature-dependent resistance at various displacement fields.** (a) Carrier density-dependent resistance at different electrical displacement field ($\bar{D}$). Temperature $T$ = 100 K (b) The same measurements as panel a except at $T$ = 37 K. Resistance goes beyond the range of our measurement system at higher displacement field. (c) Fitting curve with thermal activation (TA) model at $\bar{D}$ = 1.63 V nm$^{-1}$. Resistances in the range of 100 - 200 K are used to extract a band gap of 140.6 meV. The resistance at each $T$ is measured at the charge neutrality point, so the resistance here is maximum value at each $T$. (d) Resistance and TA fitting at different $\bar{D}$. Logarithmic y axis was used to fully show the agreement between data and fitting curves. (e) Arrhenius plot of resistance with y axis as natural logarithm of resistance and x axis as 1/$T$. Meshed area labels abnormal data due to the limited range of our measurement system as also shown in panel b.

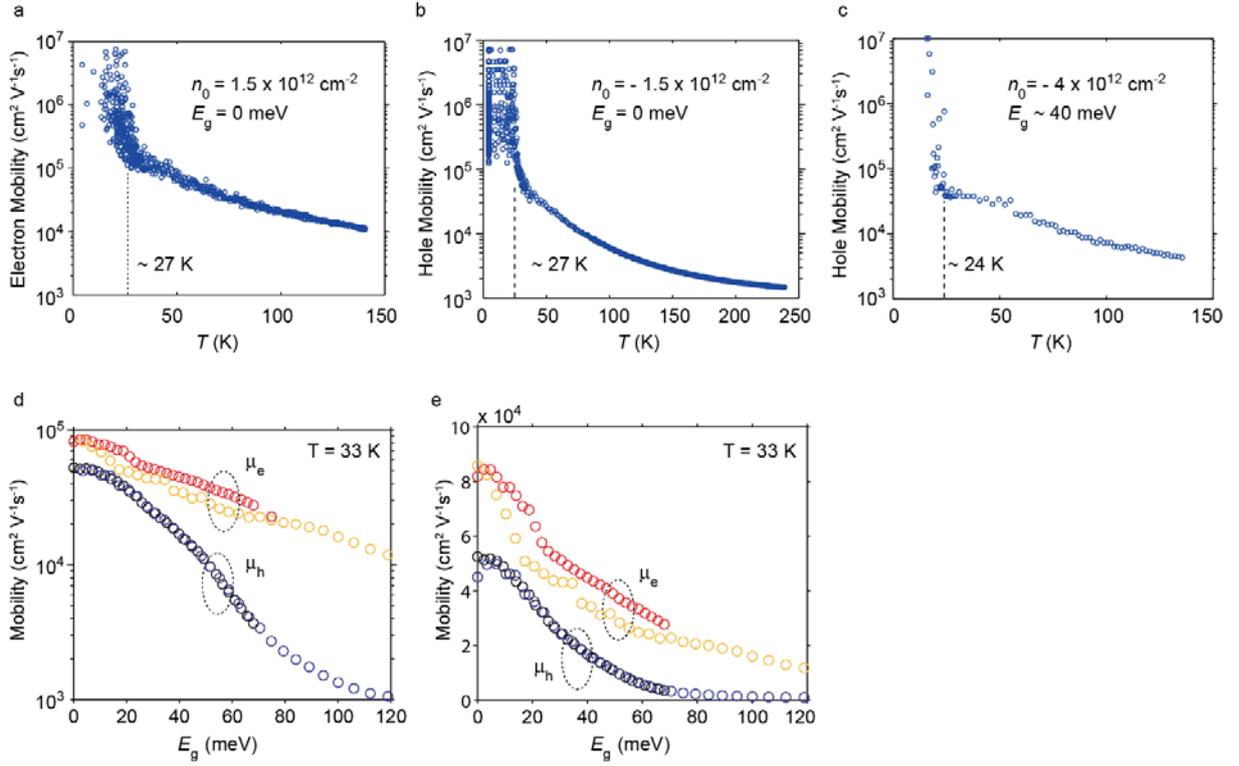

**Fig. S6. Mobility ($\mu$) of gapped bilayer graphene.** Van der Pauw method is used to measure temperature (T)-dependent carrier mobility of bilayer graphene with (a) zero gap $E_g = 0$ meV and gate-induced electron density $n_0 = 1.5 \times 10^{12}$ cm$^{-2}$ (b) $E_g = 0$ meV and $n_0 = -1.5 \times 10^{12}$ cm$^{-2}$ (c) $E_g \sim 40$ meV and $n_0 = -4 \times 10^{12}$ cm$^{-2}$. At temperature lower than 24~27 K as indicated by dashed lines, the voltage drop across the device is very small yielding large apparent values of mobilities; these features may arise from a transition to a ballistic regime. We note, however, in our analysis, we consistently work with temperatures higher than that corresponding to this regime. (d) Electron $\mu_e$ and hole $\mu_h$ mobility measured by transport measurements at $T \sim 33$ K. Red and black circles are measured with source-drain bias $V_b$ at 1 mV. Orange and blue circles are measured with $V_b$ that ranges from 1 mV to 90 mV for different $E_g$. The change of $V_b$ here is due to the measurement circuit, which has a resistor (1 MΩ) in serial in order to protect the device from burning. This is further explained in caption of Fig. S9a. (e) The same mobility data with panel e with logarithmic y axis.

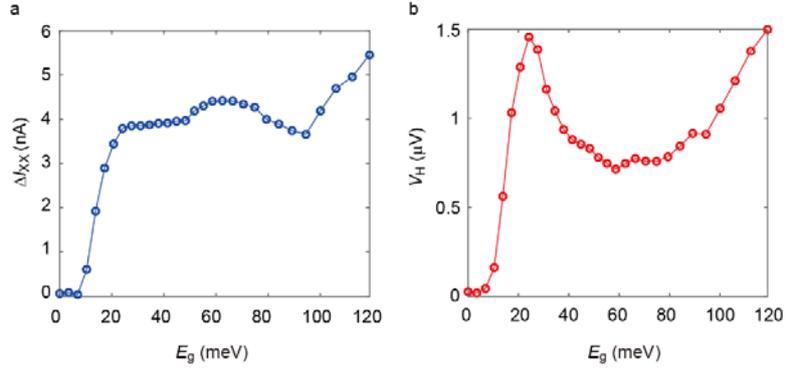

**Fig. S7. Gap-dependent valley selective Hall voltage $V_H$ and longitudinal photocurrent $\Delta I_{xx}$**
(a) and (b) Vertical linecuts of Fig. 2E ($\Delta I_{xx}$ map) and Fig. 2F ($V_H$ map) around neutrality points at $\delta D \cdot \varepsilon_0 e^{-1} \approx 2.8 \times 10^{10}$ cm$^{-2}$.

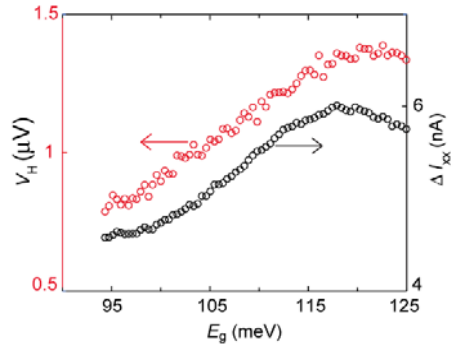

**Fig. S8. Valley selective Hall voltage $V_H$ and longitudinal photocurrent $\Delta I_{xx}$** at resonant absorption. They both show maximum values at resonance absorption. Parameters are $V_b = 50$ mV, $h\nu = 122.9$ meV, $P = 1.4$ μW μm$^{-2}$.

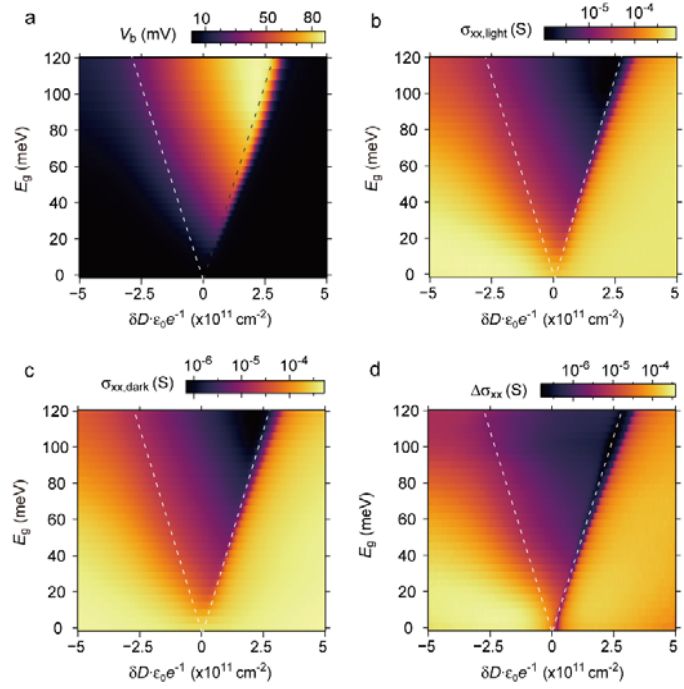

**Fig. S9. Source-drain voltage $V_b$ and Longitudinal conductivity $\sigma_{xx}$.** (a) Distribution of $V_b$ among states with different $\delta D$ and $E_g$ values, in which $x$ axis is $\delta D \cdot \varepsilon_0 e^{-1}$ ($\varepsilon_0$ and $e$ are vacuum permittivity and elementary charge). $V_b$ changes with $E_g$ and $\delta D \cdot \varepsilon_0 e^{-1}$ because a resistor (1 MΩ) is connected in series with the device, which protects the device from burning by relieving the voltage drop at the device when gapped bilayer graphene is doped (conductive). $V_b$ shows higher value in the triangle area, where the Fermi level is inside the gap and the gapped bilayer graphene in insulating. The axes are explained in Method. (b) Longitudinal conductivity $\sigma_{xx,light}$ with laser ($h\nu = 122.9$ meV, $P = 1.4$ μW μm$^{-2}$) on. (c) Longitudinal conductivity $\sigma_{xx,dark}$ without illumination. (d) Longitudinal photoconductivity $\Delta\sigma_{xx} = \sigma_{xx,light} - \sigma_{xx,dark}$. The photoconductivity shows higher values outside the gap.

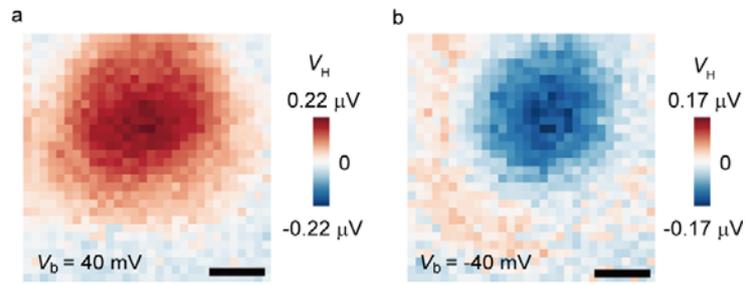

**Fig. S10. Opposite valley selective Hall voltage ($V_H$) at opposite source-drain bias ($V_b$).** (a) $V_b$= 40 mV. (b) $V_b$ = - 40 mV. Based on equation 1, the polarity of source-drain bias is directly linked to the polarity of valley selective Hall voltage. The data is measured at $E_g$= 35 meV. The scale bars indicate 10 μm.

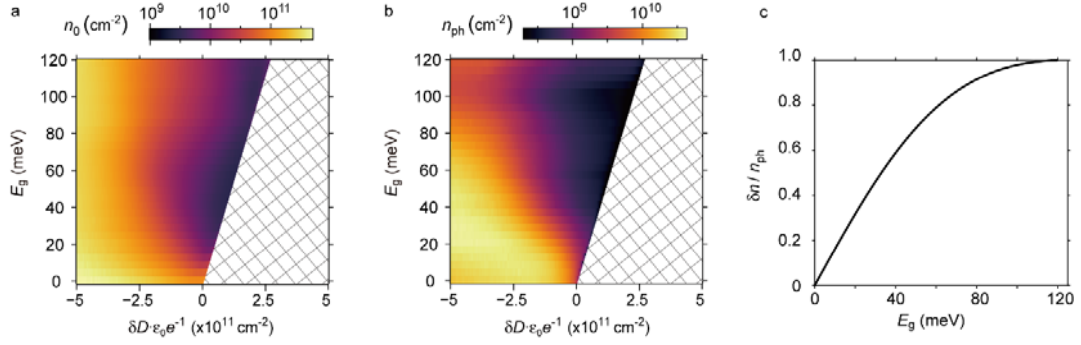

**Fig. S11. Initial carrier density before excitation ($n_0$) and photocarrier density ($n_{ph}$).** (a) Distribution of $n_0$ among states with different $\delta D$ and $E_g$ values. The x axis is $\delta D \cdot \varepsilon_0 e^{-1}$. At $V_b = 0$ and $\overline{D} = 0$ ($E_g = 0$), $\delta D \cdot \varepsilon_0 e^{-1}$ precisely equals to gate-injected carrier $n_0$, whereas at $\overline{D} \neq 0$ ($E_g \neq 0$), $\delta D \cdot \varepsilon_0 e^{-1}$ becomes smaller than $n_0$, because part of $\delta D \cdot \varepsilon_0 e^{-1}$ is consumed for moving Fermi level out of band gap. With the presence of source drain bias ($V_b \neq 0$), especially $V_b$ is comparable with $E_g$, it introduces additional charge carriers and elevate conductivity. The whole $n_0$, including gate and bias induced charge carriers, is calibrated by using $\sigma_{xx} = n_0 e \mu$, in which $\mu$ is hole mobility as calculated in Fig. S6. Here, only $n_0$ values at hole side (trapezoid) is calibrated and then used for calculating valley selective Hall conductivity $\sigma_H$ in Fig. 4 in main text. (b) Distribution of $n_{ph}$ among states with different $\delta D$ and $E_g$ values. For each $E_g$, $n_{ph}$ is calculated by dividing $\Delta \sigma_{xx}$ in Fig. S9d with average mobility $\mu_{ave} = (\mu_e + \mu_h)/2$ ($\mu_e$ and $\mu_h$ are orange and blue circles in Fig. S6d), as photocarriers include both electrons and holes. (c) The gap-dependent photocarrier imbalance ($\delta n$) between K and K' valleys that is used in the theoretical models in Fig. 4. The imbalance is caused by photoexcitation from circularly-polarized laser due to the valley selection rule, and defined as $\delta n = \left| n_{ph}^K - n_{ph}^{K'} \right|$, in which $n_{ph}^K$ ($n_{ph}^{K'}$) is photocarrier densities at K (K') valley and $n_{ph} = n_{ph}^K + n_{ph}^{K'}$. This dependence is based on the simple estimate as shown in equation S15.

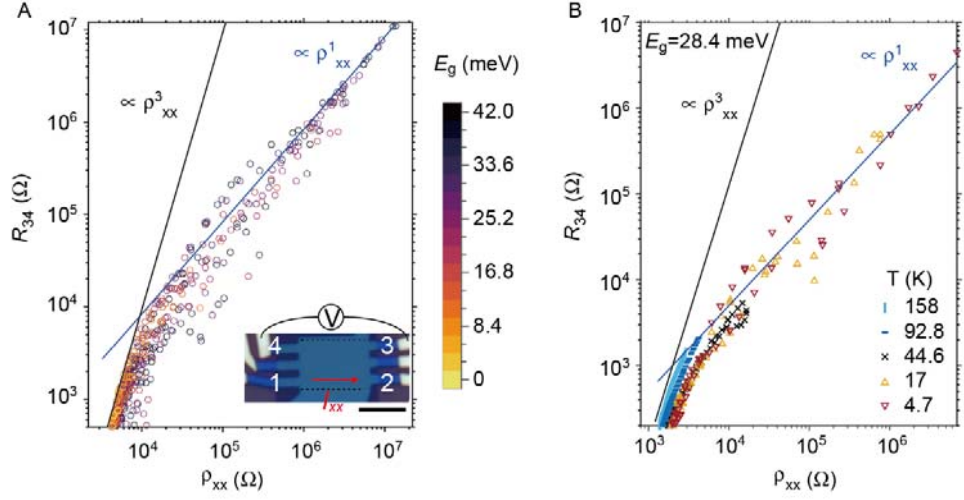

**Fig. S12. Van der pauw (vdPw) measurement of device 2** (A) vdPw resistance $R_{34}$ vs. longitudinal resistivity $\rho_{xx}$ at different band gaps $E_g$. The vdPw resistance is defined as $R_{34} = V_{34}/I_{xx}$, in which $V_{34}$ is voltage between terminal 3 and 4, and $I_{xx}$ is current injected from terminal 1 to 2, as shown in an inset figure. $\rho_{xx}$ is calculated as $\rho_{xx} = \alpha * V_{12}/I_{xx}$, where $\alpha$ is estimated as 0.157 based on a diffusive model (*39*). Blue and black lines indicate scaling relations of $R_{34} \propto \rho^1_{xx}$ and $R_{34} \propto \rho^3_{xx}$. The scale bar of inset figure indicates 5 μm. The dashed lines in inset figure indicate bilayer graphene channel region. (B) $R_{34}$ vs. $\rho_{xx}$ at different temperatures while $E_g$ is fixed at 28.4 meV. Blue and black lines indicate scaling relations of $R_{34} \propto \rho^1_{xx}$ and $R_{34} \propto \rho^3_{xx}$.